\documentclass[nohyper,twoside]{JHEP3}
\usepackage{amscd}
\usepackage{amssymb}
\usepackage{array}

\title{On the coupling of tensors to gauge fields: \\$D=5$, $N=2$
  supergravity revisited}

\author{Laura Andrianopoli\\ Centro E. Fermi,\\ Compendio Viminale,
  I-00184 Roma, and\\ Istituto Nazionale di Fisica Nucleare (INFN)\\
  Sezione di Torino, Italy
  \\ E-mail: \email{laura.andrianopoli@cern.ch}
  }
\author{Riccardo D'Auria\\ Dipartimento di Fisica,
  Politecnico di Torino,\\ Corso Duca degli Abruzzi 24, I-10129
  Torino, Italy and\\ Istituto Nazionale di Fisica Nucleare (INFN)\\
  Sezione di Torino, Italy
  \\ E-mail:  \email{riccardo.dauria@polito.it},
                            }
\author{Luca Sommovigo\\
Departamento de F\'isica Te\`orica,
Universidad de Valencia\\
C/Dr. Moliner, 50
46100 Burjassot, Valencia,
Spain
\\ E-mail: \email{luca.sommovigo@uv.es}
 }


\preprint{}

\abstract{A general free differential algebra encoding the anti-Higgs
  mechanism among two-index antisymmetric tensors and gauge vectors is
  analyzed at the full group theoretical level.    $N=2$
  supergravity in five dimensions coupled to tensor, vector and hyper
  multiplets with all possible couplings included is reconsidered from this point of
  view.  Within our approach, we find that some of the constraints on the couplings usually considered are too stringent and may in fact be relaxed. This generalization  also affects the scalar potential.}

\begin{document}

\newcommand{\rtr}{\mathrm{tr}}
\newcommand{\rU}{\mathrm{U}}
\newcommand{\rUSp}{\mathrm{USp}}
\newcommand{\rSU}{\mathrm{SU}}
\newcommand{\rE}{\mathrm{E}}
\newcommand{\rSO}{\mathrm{SO}}
\newcommand{\rSL}{\mathrm{SL}}
\newcommand{\cV}{\mathcal{V}}
\newcommand{\fsl}{\mathfrak{sl}}
\newcommand{\fgl}{\mathfrak{gl}}
\newcommand{\fe}{\mathfrak{e}}
\newcommand{\fo}{\mathfrak{o}}
\newcommand{\iden}{\mbox{{1}\hspace{-.11cm}{l}}}
\newcommand{\nn}{\nonumber}
\newcommand{\mm}{\hfill\cr}
\newcommand{\noi}{\noindent}

\def\R{\mathbb{R}}
\def\C{\mathbb{C}}
\def\Z{\mathbb{Z}}
\def\Hb{\mathbb{H}}
\def\cM{\mathcal{M}}
\def\cI{\mathcal{I}}
\def\cL{\mathcal{L}}
\def\cK{\mathcal{K}}
\def\cF{\mathcal{F}}
\def\cN{\mathcal{N}}
\def\cH{\mathcal{H}}
\def\cR{\mathcal{R}}
\def\eq#1{(\ref{#1})}
\def\IC{\relax\,\hbox{$\inbar\kern-.3em{\rm C}$}}
\def\inbar{\vrule height1.5ex width.4pt depth0pt}
\def\tI{{\tilde I}}
\def\tJ{{\tilde J}}
\def\tK{{\tilde K}}
\def\tL{{\tilde L}}
\def\tM{{\tilde M}}
\def\tN{{\tilde N}}
\def\a{\alpha}
\def\ap{\alpha'}
\def\b{\beta}
\def\bfone{\relax{\rm 1 \kern-.35em 1}}
\def\bfnull{\relax{\rm O \kern-.635em 0}}
\def\bos{{\rm bos}}
\def\c{\chi}
\def\cb{\bar{\chi}}
\def\cc{{\rm c.c.}}
\def\cov{\mathcal{D}}
\def\d{\delta}
\def\D{\Delta}
\def\de{{\rm d}\hskip -1pt}
\def\di{\displaystyle}
\def\der{\partial}
\def\dx{\right}
\def\e{\epsilon}
\def\f{\varphi}
\def\g{\gamma}
\def\G{\Gamma}
\def\i{{\rm i}}
\def\ib{{\bar{\imath}}}
\def\im{{\rm Im}\op{N}}
\def\imez{\frac{{\rm i}}{2}}
\def\j{\jmath}
\def\jb{{\bar{\jmath}}}
\def\k{\kappa}
\def\l{\lambda}
\def\L{\Lambda}
\def\lag{{\mathcal{L}}}
\def\m{\mu}
\def\mez{\frac{1}{2}}
\def\n{\nu}
\def\M{\mathcal{M}}
\def\N{\mathcal{N}}
\def\U{\mathcal{U}}
\def\na{\nabla}
\def\o{\omega}
\def\O{\Omega}
\def\ol{\overline}
\def\ot{\otimes}
\def\p{\psi}
\def\qu{\frac{1}{4}}
\def\r{\rho}
\def\re{{\rm Re}\op{N}}
\def\s{\sigma}
\def\S{\Sigma}
\def\sx{\left}
\def\t{\tau}
\def\th{\theta}
\def\Th{\Theta}
\def\ul{\underline}
\def\ve{\varepsilon}
\def\z{\zeta}

\section{Introduction}
\label{intro}
The role of tensor multiplets in supergravity has seen in the last
years a revived interest, in connection with the study of flux
compactifications in superstring or M-theory.

Two-index antisymmetric tensors are 2-form gauge fields whose
field-strengths are invariant under the (tensor)-gauge transformation
$B \to B + \de \L$, $\Lambda$ being any 1-form. A physical pattern to
introduce  massive tensor fields is the anti-Higgs mechanism, where
the dynamics allows the tensor  to take  a mass  by a suitable
coupling to some  vector field. The mass term plays the role of
magnetic charge in the theory.
The investigation of the role of massive tensor fields  was
particularly fruitful for the $N=2$ theory in 4 dimensions, where the
study of the coupling of {\em tensor-scalar} multiplets (obtained by
Hodge-dualizing  scalars covered by derivatives in the hypermultiplet
sector) to $N=2$ supergravity was considered, both as a CY compactification \cite{Louis:2002ny} and at a purely four dimensional supergravity level \cite{deWit:1982na,Theis:2003jj}. When this model was extended, in \cite{D'Auria:2004yi,Dall'Agata:2003yr,Sommovigo:2004vj}, to include the coupling to gauge multiplets, it  allowed to construct new gaugings
containing also magnetic charges, and
to find the electric/magnetic duality completion of the $N=2$ scalar
potential.

However, a general formulation of  $N=2$, $D=4$ supergravity  coupled
to {\em tensor-vector } multiplets (obtained by Hodge-dualizing
scalars in the vector multiplet sector) is still missing, even if
important steps in that direction appeared quite recently
\cite{Gunaydin:2005df,Gunaydin:2005bf}.
On the other hand, the situation appears more promising in five
dimensional supergravity. There,  2-index antisymmetric tensors appear in the gauge
sector, since the field-strengths of massless two-index tensors are
Hodge-dual to vector field-strengths, and they naturally appear in the compactification
of higher dimensional theories \footnote{For
  example, the $N=8$ gauged theory requires that a subset of the gauge
  vectors be dualized to tensors
  \cite{Gunaydin:1984qu,Gunaydin:1985cu,Pernici:1985ju}.}.
Various approaches to construct a general
 coupling to tensor multiplets in the $N=2$ theory have been  given
 \cite{Gunaydin:1983rk,Gunaydin:1983bi,Gunaydin:1984pf,Sierra:1985ax,Lukas:1998yy,Gunaydin:1999zx,Ceresole:2000jd,Bergshoeff:2004kh,Gunaydin:2005bf}.
Towards a general understanding of the four dimensional case, we adopted the strategy of first
looking at the five dimensional theory in a framework as general as possible. In particular,
an ingredient generally used for the construction of the couplings is the  ``self-duality in
odd dimensions'' \cite{Townsend:1983xs} that allows to work with massive, self-dual tensors
from the very beginning. However, in this way much of the algebraic structure underlying the
theory is not manifest. To find the most general theory in five dimensions in a way which can
give insight into the algebraic structure  also for the four dimensional case,  we found
useful to examine  first at the bosonic level and in full generality the algebraic structure which any theory coupled to tensors and gauge vectors is based on. This requires the extension
of the notion of gauge algebra to that of free differential algebra (FDA in the following)
that naturally accomodates in a general algebraic structure the presence of $p$-forms ($p>1$).
We have then devoted the first part of the paper,  section \ref{generalities}, to the study of
the gauge properties of a general FDA involving gauge vectors (1-forms) and two-index
antisymmetric tensors (2-forms). The discussion will be completely general, and will not rely
on the dimensions of space-time (apart from the obvious request $D\geq 4$, in order to have
dynamical 2-forms) nor on supersymmetry. Our procedure allows the FDA structure to be further
generalized, for $D\geq 5$, by including also couplings to higher order forms, as is the case,
in general, for flux compactifications. This is left to a future investigation.

When applying our results to the case of $D=5$, $N=2$ supergravity, in
section \ref{susy}, we
find some possible generalizations with respect to the current
literature in the subject
\cite{Gunaydin:1983rk,Gunaydin:1983bi,Gunaydin:1984pf,Sierra:1985ax,Lukas:1998yy,Gunaydin:1999zx,Ceresole:2000jd,Bergshoeff:2004kh,Gunaydin:2005bf}.
Besides the fact, already pointed out in \cite{Bergshoeff:2004kh} and
\cite{deWit:2004nw}, that it  is possible to include in the 3-form
field-strength a coupling of the type $d_{\L\S M}F^\L \wedge A^\S$
(where $\L$ enumerates  gauge fields and $M$ tensor fields), we find
that the mass matrix for the tensor fields, which in five dimensional
supergravity has to be antisymmetric ($m^{MN}=-m^{NM}$), is however
not necessarily proportional to the symplectic metric
$\Omega^{MN}=\pmatrix{0& \bfone\cr -\bfone &0}$, which is the case to
which the literature on the subject usually refers to. On the
contrary, any general antisymmetric matrix may be considered. This may
be understood, for example, by looking at the  $D=5$ $N=2$ theory
obtained by Scherk--Schwarz dimensional reduction from six dimensions
\cite{Andrianopoli:2004xu}. In this case, indeed, the tensor
mass-matrix is the Scherk--Schwarz phase and has in general different
eigenvalues. Therefore for a  general five dimensional $N=2$ theory
the generators of the gauge algebra are not necessarily in a
symplectic representation, and constraints from supersymmetry give
milder constraints on the gauging  than the ones usually considered.
As a consequence of these generalizations, the scalar potential of the
theory has some differences with respect to the previous investigations.

The FDA approach allows to interpret the resulting structure in a
general group-theoretical way which is not evident with other
approaches. Our starting point is a general  gauge algebra, which is
represented via generators with indices in the adjoint representation
of the gauge group. The building blocks of the FDA are then p-form
{\em potentials} (with, for our case, p = 1,2, that is
$A=A_\mu \de x^\mu$ and $B=B_{\mu\nu} \de x^\mu \wedge \de x^\nu$) and
their field-strengths.
Let us emphasize that in this way the fields are subject to {\em gauge
  constraints}, and are therefore {\em massless}.
The mechanism for  which the 2-forms become massive is left to the
dynamics of the Lagrangian (or alternatively, in the supersymmetric
case, also of the supersymmetric Bianchi identities). At the bosonic
level, this is implemented via the anti-Higgs mechanism, that is by
fixing the gauge invariance of the system $(A,B)$:
\begin{eqnarray}
\left\{\matrix{\delta B &=& \de \Lambda \hfill \cr
\delta A &=& \de \Theta -m \Lambda}\right.,
\end{eqnarray}
with field-strengths
\begin{eqnarray}
\left\{\matrix{H &=& \de B \hfill \cr
F &=& \de A +m B}\right.,
\end{eqnarray}
via the tensor-gauge fixing $\bar \Lambda = \frac 1m A$.

Since our analysis does not rely on the space-time dimension, we
expect to retrieve in particular,  with our approach, also the results
already known for the $D=5$, $N=2$ theory. However, there is a subtle
point here, because it appears not evident how to reconcile  the
anti-Higgs mechanism with  the fact that supersymmetry constrains
massive tensors in $D=5$ supergravity to obey the self-duality
condition:
\begin{equation}
m\partial_{[\mu }B_{\nu\rho]}\propto
\epsilon_{\mu\nu\rho\sigma\lambda}B^{\sigma\lambda}\,,\qquad \mu ,\nu,\cdots =
0,1,\dots,4.
\end{equation}
The way out from this puzzle may be found by looking again at the
subclass of models obtained by Scherk--Schwarz dimensional reduction
from six dimensions. Indeed, the six-dimensional Lorentz algebra
admits as irreducible representations self-dual tensors, satisfying
\begin{equation}
\partial_{[\hat\mu} B_{\hat\nu\hat\rho]M} = \frac{1}{6}
\epsilon_{\hat\mu \hat\nu \hat\rho \hat\sigma \hat\lambda \hat\tau}
\partial^{\hat\sigma} B^{\hat\lambda \hat\tau}_M\,,\qquad \hat\mu ,\hat\nu,\cdots =
0,1,\dots,5 .\label{sd6}
\end{equation}
Since $N=2$ matter-coupled supergravity in six dimensions contains one antiself-dual and $n_T$
self-dual tensors in the vector representation of $SO(1,n_T)$, one can use the $SO(n_T)\subset
SO(1,n_T)$ global symmetry of the model to dimensionally reduce the theory on a circle down to
five dimensions \`a la Scherk--Schwarz \cite{Andrianopoli:2004xu}, with S-S phase $m^{MN}= -
m^{NM} \in SO(n_T)$:
\begin{equation}
B_{\hat\mu\hat\nu M} (x,y_5)=\Bigl({\exp}[m y_5]\Bigr)_M^{\phantom{M}N} \sum_n
B^{(n)}_{\hat\mu\hat\nu N}(x)\exp\left[\frac{\i n }{2\pi
    R}y_5\right]\,.\label{ss}
\end{equation}
Applying \eq{ss} to the self-duality relation \eq{sd6}, we find
\begin{equation}
\partial_{[\mu} B_{\nu\rho]M} = \frac16
\epsilon_{\mu\nu\rho\sigma\lambda 5}\left(m_M{}^N B_N^{\sigma\lambda}
+ 2 F^{\lambda\sigma}_N\right) \,,\qquad \mu=0,1,\dots,4\label{sd5}
\end{equation}
where $F_{\lambda\sigma N}\equiv\partial_{[\sigma} B_{\lambda] 5 N}$.
Eq. \eq{sd5} expresses the self-duality obeyed by the tensors in five
dimensional supergravity. However, it also shows that the
field-strengths of the vectors $B_{\mu 5 N}$, that give mass to the
tensors $B_{\mu\nu M} $ via the anti-Higgs mechanism, are in fact the
Hodge-dual of the tensors $B_{\mu\nu M} $ themselves.
From our analysis applied to $N=2$ supergravity in five dimensions, we
find this  to be a general fact, not necessarily related to theories admitting
 a six dimensional uplift: in each case, the massive tensor fields
belong to short representations of supersymmetry, and the dynamical
interpretation of the mechanism  giving mass to the tensors requires
the coupling of the massless tensors to gauge vectors which are the
Hodge-dual of the tensors themselves.

The paper is organized as follows: In section \ref{generalities} we
study the general FDA describing the coupling of two-index antisymmetric
tensor fields to non-abelian gauge vectors and show in detail, for the general case, how the anti-Higgs mechanism takes place. In section
\ref{susy}, we apply the formalism to the case of $N=2$ five
dimensional supergravity, using the geometric approach to find the
Lagrangian,  supersymmetry transformations rules and constraints on
the scalar geometry and gauging. Our results are summarized in the
concluding section, while we left to the appendices some technical
details and the comparison of our  notations with the ones of
\cite{Gunaydin:1983bi} and of \cite{Ceresole:2000jd}.


\section{A general bosonic  theory with massive tensors and
  non-abelian vectors}
\label{generalities}

In this section we are going to study the gauge structure of a general
theory with two-index antisymmetric tensor fields coupled to
gauge vectors. The discussion here will be general, with no need to
make reference to any particular dimension of space-time nor to any
possible supersymmetric extension of the model. Later, in section
\ref{susy}, we will consider the supersymmetrization of the model,
specifying the discussion to the case of $N=2$ five dimensional
supergravity coupled to vector, tensor and hyper multiplets. The corresponding four
dimensional case of $N=2$ supergravity coupled to vector-tensor
multiplets is under investigation, and is left to a future publication.
\subsection{FDA and the anti-Higgs mechanism}
\subsubsection{Abelian case}
The simplest case of a FDA including 1-form and 2-form potentials \footnote{0-forms will also
be included in section \ref{susy}, when considering a supersymmetric version of the theory}
is described by a set of abelian gauge vectors $A^M$ and of massless tensor two-forms $B_M$
($M=1,\dots n_T$.) interacting  by a coupling $m^{MN}$. The field-strengths are:
\begin{eqnarray}
\left\{
\matrix{F^M&=& \de A^M + m^{MN} B_N\hfill\cr
H_M &=& \de B_M \hfill }\right.
\end{eqnarray}
and are invariant under the gauge transformations:
\begin{eqnarray}
\left\{
\matrix{\delta A^M &=& \de \Theta^M- m^{MN} \L_N \mm
\delta B_M &=& \de \L_M \hfill}\right.
\end{eqnarray}
 with  $\Theta^M$ parameters of
infinitesimal U(1) gauge transformations and $\L_M$ one-form
parameters of infinitesimal tensor-gauge transformations of the
two-forms $B_M$. In this case the system undergoes the anti-Higgs
mechanism, and it is possible to fix the tensor-gauge so that:
\begin{eqnarray}
\left\{\matrix{A^M &\to & A'^M= - m^{MN} \bar\L_N\mm
B_M & \to &  B'_M = B_M + \de \bar \L_M;} \right.
\end{eqnarray}
In this way the gauge vectors $A^M$ disappear from the spectrum
providing the degrees of freedom necessary for the tensors to
acquire a mass, since:
\begin{eqnarray}
\left\{\matrix{ F'^M &=&  m^{MN} B_N\mm
H'_M & = & \de B_M.\hfill}\right.\label{antihiggs0}
\end{eqnarray}

\subsubsection{Coupling to a non-abelian algebra}
The model outlined above may be generalized by including the  coupling of this
system to $n_V$ gauge vectors $A^\L$ ($\L = 1, \dots n_V$), with
gauge algebra $G_0$ (not necessarily semisimple), if the index $M$ of
the tensors $B_M$ and of the abelian vectors $A^M$ runs over a
representation of $G_0$. In this case the FDA becomes \footnote{We will generally assume, here and in the following, that the tensor mass-matrix $m^{MN}$ is invertible. In case it has some 0-eigenvalues, we will restrict to the submatrix with non-vanishing rank. This is not a restrictive assumption, because any tensor corresponding to a zero-eigenvalue of $m$ may be dualized to a gauge vector and so included in the set of $\{A^\L\}$.}:
\begin{eqnarray}
\left\{\matrix{F^\L &=& \de A^\L + \mez f_{\S\G}{}^\L  A^\S \wedge
  A^\G \mm
F^M &=& \de A^M - T_{\L N}{}^M A^\L \wedge A^N + m^{MN} B_N \mm
& \equiv & D A^M + m^{MN} B_N \mm
H_M &=& \de B_M + T_{\L M}{}^N A^\L \wedge B_N + d_{\L NM}  F^\L\wedge
  A^N  \mm
&\equiv & D B_M + d_{\L NM}  F^\L\wedge A^N \hfill}
  \right. \label{couplings}
\end{eqnarray}
Here $f_{\S\G}{}^\L$ are the structure constants of the gauge algebra
$G_0$ and $T_{\L M}{}^N $, $d_{\L MN}$ suitable couplings. The
closure of the FDA (${\de}^2 A^\L = {\de}^2 A^M = {\de}^2 B_M = 0$)
gives the following constraints:
\begin{eqnarray}
f_{[\L\S}{}^\D f_{\G]\D}{}^\O&=&0 \label{gaugealgebra1}  \\
T_{[\Lambda | M}{}^P T_{\Sigma ] P}{}^N & = &\frac 12  f_{\L \S}{}^\G T_{\G M}{}^N\label{gaugealgebra2}  \\
T_{\L M}{}^N &=& -d_{\L MP }m^{NP} =d_{\L PM}m^{PN}\label{gaugealgebra3} \\
T_{\L N}{}^M m^{NP} &=& -T_{\L N}{}^P m^{MN}\label{gaugealgebra4}  \\
T_{\S M}{}^N d_{\G PN}& +& T_{\S P}{}^N d_{\G NM} - f_{\S\G}{}^\L
d_{\L P M}{}^Q = 0. \label{gaugealgebra5}
\end{eqnarray}
Eq.s \eq{gaugealgebra1}, \eq{gaugealgebra2} show in particular that
the structure constants $f_{\L\S}{}^\G$
do indeed close the algebra $G_0$ and that $T_{\L M}{}^N$ are
generators of $G_0$  in the representation spanned by the tensor
fields. Eq.s \eq{gaugealgebra3} and \eq{gaugealgebra4} imply:
\begin{equation}
\matrix{m^{MN} &=& \mp m^{NM}\mm
d_{\L MN} &=& \pm d_{\L NM},\hfill}
\end{equation}
and \eq{gaugealgebra5} is a consistency condition that, when multiplied by
$m^{PQ}$, is equivalent to \eq{gaugealgebra2} (upon use of \eq{gaugealgebra4}).

When \eq{gaugealgebra1} - \eq{gaugealgebra5} are satisfied, the Bianchi identities read:
\begin{eqnarray}
\left\{\matrix{\de F^\L + f_{\S\G}{}^\L  A^\S \wedge F^\G &=& 0 \mm
\de F^M -  T_{\L N}{}^M A^\L \wedge F^N &=& m^{MN} H_N \mm
\de H_M +  T_{\L M}{}^N A^\L \wedge H_N &=& d_{\L MN} F^N \wedge F^\L.
 \hfill}\right.  \label{bisimplified}
\end{eqnarray}
To see how the anti-Higgs mechanism works in this more general case,
let us give the gauge  and tensor-gauge transformations  of the fields
(including the non-abelian transformations belonging to $G_0$, with parameter
$\e^\L$). They become:
\begin{eqnarray}
\left\{\matrix{\delta A^\L &=& \de \e^\L + f_{\S \G}{}^\L  A^\S \e^\G
  \equiv D \e^\L \mm
\delta A^M &=& \de \Theta^M - T_{\L N}{}^M A^\L \Theta^N + T_{\L
  N}{}^M A^N \e^\L  - m^{MN} \L_N \mm
& \equiv & D \Theta^M + T_{\L N}{}^M A^N \e^\L  - m^{MN} \L_N \mm
\delta B_M &=& \de \L_M + T_{\L M}{}^N A^\L \wedge \L_N - d_{\L MN}
  F^\L \Theta^N - T_{\L M}{}^N B_N \e^\L \mm
& \equiv & D \L_M - d_{\L MN} A^\L \wedge \de \Theta^N - T_{\L M}{}^N
  B_N  \e^\L ,\hfill} \right. \label{gaugeinvar1}
\end{eqnarray}
with:
\begin{eqnarray}
\left\{\matrix{\delta F^\L &=& f_{\S \G} {}^\L F^\S \e^\G \mm
\delta F^M &=& T_{\L N}{}^M F^N   \e^\L \mm
\delta H_M &=& - T_{\L M}{}^N H_N \e^\L.\hfill}\right.
\end{eqnarray}
Fixing the gauge of the tensor-gauge transformation as:
\begin{eqnarray}
\left\{\matrix{ A^\L &\to& A'^\L =A^\L \mm A^M &\to& A'^M= - m^{MN} \bar \L_N \mm B_M &\to&
B'_M = B_M + D \bar \L_M,\hfill}\right. \label{gaugefix1}
\end{eqnarray}
we find:
\begin{eqnarray}
\left\{\matrix{F'^\L &=& F^\L \mm
F'^M &=& m^{MN} B_N \mm
H'_M &=& D B_M \hfill}\right. \label{gaugefixed1}
\end{eqnarray}
When the tensor-gauge is fixed as in \eq{gaugefix1},\eq{gaugefixed1},
the vectors $A^M$ disappear from the spectrum while the tensors $B_M$
acquire a mass. As anticipated in the introduction, this is in
particular the starting point of the  formulation adopted in the literature to describe
$D=5$, $N=2$ supergravity coupled to massive tensor multiplets
\cite{Gunaydin:1999zx,Ceresole:2000jd,Bergshoeff:2004kh}.

\bigskip

However, let us observe that in this more general case the abelian
gauge vectors $A^M$, providing the degrees of freedom needed to give a
mass to the tensors via the anti-Higgs mechanism, are charged under
the gauge algebra $G_0$. It is not possible to make the gauge
transformation of the vectors $A^M$ compatible with that of the $A^\L$
unless all together the vectors $\{A^\L , A^M\} \equiv A^{\tilde I} $
form the co-adjoint representation of some larger non semisimple gauge
algebra $G\supset G_0$.

The relations so far obtained may then be written with the collective
index ${\tilde I} =(\L , M)$, in terms of  structure constants
$f_{{\tilde J}{\tilde K}}{}^{\tilde I}$ restricted to the following
non vanishing entries: 
\begin{equation}
f_{{\tilde J}{\tilde K}}{}^{\tilde I}= ( f_{\L\S}{}^\G ,  f_{\L
  M}{}^N=-T_{\L M}{}^N)\,,
\label{algebrasimplified}
\end{equation}
and of the couplings:
\begin{eqnarray}
m^{\tilde I M}\equiv \delta^{\tilde I}_N m^{NM}\,, \qquad d_{\tilde I
  \tilde J M} \equiv  \delta_{\tilde I}^\L \delta_{\tilde J}^N d_{\L
  NM}.\label{restrictions}
\end{eqnarray}
In terms of the tilded quantities the FDA \eq{couplings} reads:
\begin{eqnarray}
\left\{\matrix{F^{\tilde I} &\equiv& \de A^{\tilde I} + \mez
  f_{{\tilde J}{\tilde K}}{}^{\tilde I} A^{\tilde J} \wedge A^{\tilde
    K} + m^{{\tilde I} M} B_M \mm
H_M &\equiv& \de B_M + T_{{\tilde I} M}{}^N A^{\tilde I} B_N +
  d_{{\tilde I}{\tilde J} M} F^{\tilde I} \wedge A^{\tilde
  J}}\right. \label{fda}
\end{eqnarray}
 with Bianchi identities:
\begin{eqnarray}
\left\{
\matrix{
\de F^{\tilde I} + \sx( f_{{\tilde J}{\tilde K}}{}^{\tilde I} +
m^{{\tilde I} M} d_{{\tilde K}{\tilde J} M} \dx)A^{\tilde J} F^{\tilde
  K}  &=  \; m^{{\tilde I} M} H_M \mm
\de H_M +\sx( T_{{\tilde I} M}{}^N + m^{{\tilde J} N} d_{{\tilde
    J}{\tilde I} M} \dx)A^{\tilde I} H_N & = \; d_{{\tilde I}{\tilde
    J} M} F^{\tilde I} F^{\tilde J} \hfill}\right. , \label{BI}
\end{eqnarray}
provided the following relations, equivalent to \eq{gaugealgebra1} - \eq{gaugealgebra5},  hold:
\begin{eqnarray}
\matrix{f_{[{\tilde I}{\tilde J}}{}^{\tilde L} f_{{\tilde K}]{\tilde
      L}}{}^{\tilde M} &=& 0 \mm
\sx[ T_{\tilde I} , T_{\tilde J} \dx] &=& f_{{\tilde I}{\tilde
      J}}{}^{\tilde K} T_{\tilde K} \mm
T_{{\tilde I} M}{}^{(N} m^{{\tilde I} |P)} &=& 0\mm
m^{{\tilde I} N} T_{{\tilde J} N}{}^M &=& f_{{\tilde J}{\tilde
      K}}{}^{\tilde I} m^{{\tilde K} M} \mm
T_{{\tilde I} M}{}^N &=& d_{{\tilde I}{\tilde J} M} m^{{\tilde J} N}
      \mm
T_{[{\tilde I} | M}{}^N d_{{\tilde K} | {\tilde J}] N} &-& (f_{ [
      {\tilde I} | {\tilde K}}{}^{\tilde L}  + m^{{\tilde L} N}
      d_{{\tilde K} [ {\tilde I} N} ) d_{{\tilde L} | {\tilde J}] M} -
      \mez f_{{\tilde I}{\tilde J}}{}^{\tilde L} d_{{\tilde K}{\tilde
      L}  M} = 0.}
\label{closure}
\end{eqnarray}

Subject to the constraints \eq{closure}, the system is covariant under
the gauge transformations:
\begin{eqnarray}
\left\{\matrix{\delta A^{\tilde I} &=& \de \e^{\tilde I} + f_{{\tilde
      J}{\tilde K}}{}^{\tilde I} A^{\tilde J} \e^{\tilde K} -
  m^{{\tilde I} M} \L_M \mm
\delta B_M &=& \de \L_M + T_{{\tilde I} M}{}^N A^{\tilde I} \L_N -
      d_{{\tilde I}{\tilde J} M} F^{\tilde I} \e^{\tilde J} -
      T_{{\tilde I} M}{}^N \e^{\tilde I} B_N}\right.\label{gaugefin}
\end{eqnarray}
implying  the gauge transformation of the field strengths:
\begin{eqnarray}
\left\{\matrix{\delta F^{\tilde I} &=& - \sx( f_{{\tilde J}{\tilde
      K}}{}^{\tilde I} + m^{{\tilde I} M} d_{{\tilde K}{\tilde J} M}
  \dx) \e^{\tilde J} F^{\tilde K} \mm
\delta H_M &=& - \sx( T_{{\tilde I} M}{}^N + m^{{\tilde J} N}
      d_{{\tilde J}{\tilde I} M} \dx) \e^{\tilde I} H_N}\right.
\label{labfalfa}
\end{eqnarray}

\subsubsection{A general FDA}
We now observe that the restrictions on the couplings
\eq{algebrasimplified} and \eq{restrictions} have been set to exactly
reproduce eqs. \eq{couplings} while exhibiting the fact that
$A^{\tilde I}$ collectively belong to the adjoint of some algebra
$G\supset G_0$. Actually eq.s \eq{couplings} and \eq{closure} allow in
fact a more general gauge structure than the one declared in
\eq{algebrasimplified}, \eq{restrictions}. Let $T_{\tilde I}\in {\rm
  Adj}\,G$ be the gauge generators dual to $A^{\tilde I}$. For the
case of \eq{algebrasimplified}, $G$ has the semisimple structure
$G=G_0\ltimes \R^{n_T}$, and the generators $T_\L \in G_0$ may be
realized in a  block-diagonal way (with entries $T_{\L \S}{}^\G
=f_{\L\S}{}^\G$, $T_{\L M}{}^N = - f_{\L M}{}^N$) while the $T_M$ are
off-diagonal (with entries $T_{M \L}{}^N =f_{\L M}{}^N$). However, any
gauge algebra $G$ with structure constants $f_{\tilde I \tilde
  J}{}^{\tilde K}$ may in principle be considered, provided it
satisfies the constraints \eq{closure}. In the general case, to match
\eq{closure} one must also relax the restrictions on the couplings
\eq{algebrasimplified}, \eq{restrictions}, and allow for more general
$f_{\tilde I \tilde J}{}^{\tilde K}$ and $d_{\tI\tJ M}$. This includes
in particular the case 
\begin{equation}
f_{\L\S}{}^M \neq 0\,, \qquad d_{\L\S M} \neq 0 \label{flsm}
\end{equation}
which was considered in \cite{Bergshoeff:2004kh} and \cite{deWit:2004nw}. In this case, $G$
 cannot be semisimple, and $G_0$ is not a subalgebra of $G$ \footnote{We acknowledge an enlightening
discussion
  with Maria~A.~Lled\'o on this point.}. This implies that the vectors
$A^M$ do not decouple anymore at the level of gauge algebra, and this,
at first sight, would be an  obstruction to implement the anti-Higgs
mechanism. However, this apparent obstruction may be simply overcome in the FDA framework, due to the freedom of
redefining the tensor fields as  \cite{Dall'Agata:2005mj}:
\begin{equation}
B_M \to B_M  + k_{{\tilde I}{\tilde J} M} A^{\tilde I} \wedge
A^{\tilde J} ,\label{red}
\end{equation}
for any $k_{{\tilde I}{\tilde J} M} $ antisymmetric in ${\tilde
  I},{\tilde J}$. It is then possible to implement the anti-Higgs
mechanism with the tensor-gauge fixing (which includes a field
redefinition as in \eq{red}):
\begin{eqnarray}
\left\{\matrix{ A^\L &\to& A'^\L =  A^\L \mm
A^M &\to& A'^M = - m^{ MN} \bar \L_N \mm
B_M &\to& B'_M = B_M  - \frac 12 d_{\L\S M}A^\L \wedge A^\S + D \bar
\L_M\hfill}\right. \label{gaugefix}
\end{eqnarray}
This still gives:
\begin{eqnarray}
\left\{\matrix{F'^\L &=& F^\L \mm
F'^M &=&  m^{MN} B_N \mm
H'_M &=& D B_M \hfill}\right. \label{gaugefixed}
\end{eqnarray}
provided that:
\begin{equation}
m^{MN} d_{[\L\S] N} = f_{\L\S}{}^M.
\end{equation}
With this observation, we may now  analyze in full generality which
non trivial structure constants may be turned on in \eq{fda} in a way
compatible with the anti-Higgs mechanism.

First of all, it is immediate to see that if:
\begin{equation}
f_{\tilde I M}{}^\Sigma \neq 0 \,,
\end{equation}
 it is impossible to implement the anti-Higgs
mechanism, because they introduce a coupling to  the gauge vectors
$A^M$  in the field-strengths $F^\Lambda$ which is not possible to
reabsorb by any field-redefinition.

 Considering then the case:
\begin{equation}
f_{MN}{}^P \neq 0 \,, \qquad d_{MNP}\neq 0 \,.
\end{equation}
we see that $f_{MN}{}^P$ would introduce a non-abelian interactions
among the vectors $A^M$ and in particular, for the case $\Lambda =0$,
this would imply that the $A^M$ close a non-abelian gauge
algebra. This case may be treated in a way quite similar to the case
\eq{flsm}, since again we may use the freedom  in \eq{red} to absorb
the non-abelian contribution to $F^M$ in a redefinition of $B_M$. The
anti-Higgs mechanism may then be implemented via the tensor-gauge
fixing: 
\begin{eqnarray}
\left\{\matrix{
A^\L &\to& A'^\L =  A^\L \mm
A^M &\to& A'^M = - m^{ MN} \bar \L_N \mm
B_M &\to& B'_M = B_M  - \frac 12 d_{NP M}A^N \wedge A^P + D \bar
\L_M\hfill}\right. \label{gaugefix2}
\end{eqnarray}
giving, as before:
\begin{eqnarray}
\left\{\matrix{F'^\L &=& F^\L \mm
F'^M &=&  m^{MN} B_N \mm
H'_M &=& D B_M \hfill}\right. \label{gaugefixed2}
\end{eqnarray}
provided that:
\begin{equation}
m^{MQ} d_{[NP] Q} = f_{NP}{}^M. \label{dmnp}
\end{equation}
This shows that also non-abelian gauge vectors $A^M$ may be
considered, and still may  decouple from
the gauge-fixed theory by giving mass to the tensors $B_M$. For this
case, however, the constraints \eq{closure}, together with \eq{dmnp},
give the following conditions on the couplings:
\begin{equation}
\left\{\matrix{d_{MNP} &=& d_{[MNP]} \mm   m^{MN}&=&+m^{NM}\hfill}\right.
.\label{msymm}
\end{equation}
As we are going to discuss in the next section, for the $D=5$, $N=2$
theory the matrix $m^{MN}$ has to be antisymmetric, and this then
implies, for this theory, $f_{MN}{}^P =0$. We conclude that even if
the algebra \eq{fda} can have non trivial extensions with new
couplings, this is not the case for the $D=5$, $N=2$ theory we shall
be concerned with  in section \ref{susy}, so that the couplings
$f_{MN}{}^P$ and $d_{MNP}$ will be set to zero. 
\bigskip

\subsection{General properties of the FDA}
\label{fdageneralities}
A further observation concerns eq.s \eq{BI} and \eq{labfalfa}. In
these equations, as in all the relations involving the physical field
strengths $F^{\tilde I}$ and $H_M$, the following objects appear:
\begin{eqnarray}
\matrix{\hat{f}_{{\tilde J}{\tilde K}}{}^{\tilde I}  &\equiv&
  f_{{\tilde J}{\tilde K}}{}^{\tilde I} + m^{{\tilde I} M} d_{{\tilde
      K}{\tilde J} M} \, ; \mm
\hat{T}_{{\tilde I} M}{}^N &\equiv& T_{{\tilde I} M}{}^N + m^{{\tilde
  J} N} d_{{\tilde J}{\tilde I} M}= 2 d_{({\tilde I}{\tilde J})M}
  m^{{\tilde J} N}. \hfill }\label{generalgenerat}
\end{eqnarray}

The generalized couplings $\hat{f}_{{\tilde J}{\tilde K}}{}^{\tilde
  I}$  belong to a representation of the gauge algebra $G$ which is
not the adjoint, since they are not antisymmetric in the lower
  indices. In particular we find : 
\begin{eqnarray}
\matrix{\hat f_{{\tilde I}{\tilde J}}{}^{\tilde K} m^{{\tilde J} M}
  &=& \hat T_{{\tilde I} N}{}^Mm^{{\tilde J} N} \mm
\hat f_{{\tilde I}{\tilde J}}{}^{\tilde K} m^{{\tilde I} M} &=&
  0.\hfill}
\end{eqnarray}
However, the $\hat{f}_{{\tilde J}{\tilde K}}{}^{\tilde I}$ and $\hat  T_{{\tilde I} N}{}^M$ can be understood as  representations of generators $\hat f_{\tilde I}$ and $\hat
T_{\tilde I}$ that still generate the gauge algebra $G$. Indeed the following relations hold  (subject to the constraints
\eq{closure}):
\begin{eqnarray}
\matrix{\sx[ \hat{f}_{\tilde I} , \hat{f}_{\tilde J} \dx] &=& -
  f_{{\tilde I}{\tilde J}}{}^{\tilde K} \hat{f}_{\tilde K},\mm
\sx[ \hat{T}_{\tilde I} , \hat{T}_{\tilde J} \dx] &=& f_{{\tilde
  I}{\tilde J}}{}^{\tilde K} \hat{T}_{\tilde K}.\hfill}
  \label{bigalgebra}
\end{eqnarray}
The generalized couplings $\hat f$ and $\hat T$  express the
deformation of the gauge structure due to the presence of the tensor
fields. In particular, only the structure constants of $G_0$ are unchanged, corresponding to the fact that this is
the algebra  realized exactly in the interacting theory
\eq{fda} after  the anti-Higgs mechanism has taken place.
The rest of the gauge algebra $G$ is instead spontaneously
broken by the anti-Higgs mechanism (which requires, if  $f_{\L\S}{}^M\neq 0$, also a tensor redefinition, as explained in \eq{gaugefix}).
However, the entire algebra $G$
is still realized, even if in a more subtle way, as eq.s
\eq{bigalgebra} show.
From a
physical point of view, this is expected by a counting of degrees
of freedom, since the degrees of freedom required to make a
two-index tensor massive are the ones of a gauge vector connection
\footnote{Indeed, the on-shell degrees of freedom of a massless
(2-index) tensor and of a vector in $D$ dimensions are
$(D-2)(D-3)/2$ and $(D-2)$ respectively, while the ones of a massive
tensor are $(D-1)(D-2)/2= (D-2)(D-3)/2 + (D-2)$.}, so that also the vectors $A^M$, besides the $A^\L$, are expected to be massless gauge vectors.
 This algebra indeed closes provided the Jacobi
identities $f_{[{\tilde I}{\tilde J}}{}^{\tilde L}  f_{{\tilde
      K}]{\tilde L} }{}^{\tilde M}  =0 $ are satisfied. We find indeed, using \eq{bigalgebra}:
      \begin{eqnarray}
\matrix{ \left[ \left[\hat f_{[{\tilde I}},\hat f_{\tilde
      J}\right],\hat f_{{\tilde K} ]}\right]_{\tilde L}{}^{\tilde
    P}&=&-f_{[{\tilde I}{\tilde J}}{}^{\tilde N} f_{{\tilde K}]{\tilde
      N} }{}^{\tilde M}  \hat f_{{\tilde M} {\tilde L} }{}^{\tilde P}
  =0 \mm
 \left[ \left[\hat T_{[{\tilde I}},\hat T_{\tilde J}\right],\hat
      T_{{\tilde K} ]}\right]_M{}^N&=&-f_{[{\tilde I}{\tilde
      J}}{}^{\tilde M} f_{{\tilde K}]{\tilde M}}{}^{\tilde L} \hat
      T_{{\tilde L}M}{}^N  =0\hfill} \label{closed}
\end{eqnarray}

The
hatted generators $\hat f$, $\hat T$   play the role of {\em physical couplings} when the gauge structure is extended to include charged tensors. They
have then to be considered as the appropriate generators of the free
differential structure. It may be useful to recast the theory in terms of all the couplings
appearing in the Bianchi identities \eq{BI}, that is the hatted
generators and the symmetric part $d_{({\tilde I}{\tilde J})M}$ of the
Chern--Simons-like coupling $d_{{\tilde I}{\tilde J} M}$. This is done
by the field redefinition:
\begin{equation}
B_M \to \tilde B_M = B_M + \frac 12 d_{[{\tilde I}{\tilde J}]M}
A^{\tilde I}\wedge A^{\tilde J} \label{bridef}
\end{equation}
so that the FDA takes the form:
\begin{eqnarray}
\left\{\matrix{F^{\tilde I} &\equiv& \de A^{\tilde I} + \mez \hat
  f_{{\tilde J}{\tilde K}}{}^{\tilde I} A^{\tilde J} \wedge A^{\tilde
    K} + m^{{\tilde I} M}\tilde B_M \mm
H_M &\equiv& \de \tilde B_M + \frac 12 \hat T_{{\tilde I} M}{}^N
  A^{\tilde I} \tilde B_N + d_{({\tilde I}{\tilde J}) M} F^{\tilde I}
  \wedge A^{\tilde J} + \mathcal{K}_{M {\tilde I}{\tilde J}{\tilde
  K}}A^{\tilde I}\wedge A^{\tilde J} \wedge A^{\tilde
  K}\hfill}\right. \label{fda2}
\end{eqnarray}
and the constraints \eq{closure} in the new formulation read, after
introducing $\tilde f_{{\tilde I}{\tilde J}}{}^{\tilde K} \equiv \hat
f_{[{\tilde I}{\tilde J}]}{}^{\tilde K}$:
\begin{eqnarray}
\matrix{\tilde f_{[{\tilde I}{\tilde J}}{}^{\tilde M} \tilde
    f_{{\tilde K}]{\tilde M}}{}^{\tilde L} &=& 2 m^{{\tilde L} M}
  \mathcal{K}_{M[{\tilde I}{\tilde J}{\tilde K} ]} \mm
\hat T_{[{\tilde I} M}{}^N  \hat T_{{\tilde J} ]N}{}^P  &=& \tilde
    f_{{\tilde I}{\tilde J}}{}^{\tilde K} \hat T_{\tilde K} + 12
    \mathcal{K}_{M{\tilde I}{\tilde J}{\tilde K}}m^{{\tilde K} P} \mm
\hat T_{{\tilde I} M}{}^{N} m^{{\tilde I} P} &=& 0\mm
\frac{1}{2} m^{{\tilde I} N} \hat T_{{\tilde J} N}{}^M &=&\tilde
    f_{{\tilde J}{\tilde K}}{}^{\tilde I} m^{{\tilde K} M} \mm
\hat T_{{\tilde I} M}{}^N &=& 2 d_{({\tilde I}{\tilde J}) M}
    m^{{\tilde J} N} \mm
\hat T_{[{\tilde I} | M}{}^N d_{( {\tilde J}]{\tilde K} )N} &-& 2\hat
    f_{ [ {\tilde I} | {\tilde K}}{}^{\tilde L} d_{ ({\tilde
    J}]{\tilde L}) M} -\tilde f_{{\tilde I}{\tilde J}}{}^{\tilde L}
    d_{({\tilde K}{\tilde L}) M} = -6 \mathcal{K}_{M{\tilde I}{\tilde
    J}{\tilde K}} \mm
\mathcal{K}_{N [{\tilde J}{\tilde K}{\tilde L}} \hat T_{{\tilde
    I}]|M}{}^N&-&3  \mathcal{K}_{M {\tilde P} [{\tilde I}{\tilde
    J}}\tilde f_{{\tilde K}{\tilde L}]}{}^{\tilde P} =0.\hfill}
\label{closure2}
\end{eqnarray}
In eq.s \eq{fda2} and \eq{closure2} we have introduced the definition:
\begin{equation}
\mathcal{K}_{M{\tilde I}{\tilde J}{\tilde K}}= \frac 12 \hat
f_{[{\tilde I}{\tilde J}}{}^{\tilde{L}} d_{({\tilde K}] {\tilde{L}})
  M} +\frac{2}{3} d_{({\tilde{L}}[{\tilde J})M} \hat f_{{\tilde
      I}]{\tilde K}}{}^{\tilde{L}}. \label{k}
\end{equation}
that could also be found by directly studying the closure of the FDA
\eq{fda2} without referring to its derivation from \eq{fda}.

Eq. \eq{fda2}, which is expressed in terms of the physical couplings
only,  is completely equivalent to \eq{fda}. This is in fact the
formulation used in
\cite{deWit:2004nw}, for the study of $N=8$ supergravity in 5
dimensions.  However, as eq.s \eq{closure2} shows, in the formulation
\eq{fda2} the gauge structure is not completely manifest, because for
the ``structure constants'' $\tilde f_{\tilde I\tilde J}{}^{ \tilde
  K}$  the Jacobi identities fail to close.

Equation \eq{fda} (or, equivalently, \eq{fda2}) is the most general FDA
involving vectors and 2-index antisymmetric
tensors.  Any other possible deformation of \eq{fda} is indeed trivial
(unless the system is also coupled to higher order forms) as we will
show in detail in Appendix \ref{deformedfda}.
\bigskip

As a final remark, let us observe that, given the definitions
 \eq{fda}, the FDA still enjoys a  scale invariance under the transformation, with parameter $\alpha$:
\begin{eqnarray}
\left\{\matrix{m^{MN} &\to& \a \, m^{MN} \mm
B_M &\to& \frac{1}{\a} B_M \mm
d_{{\tilde I}{\tilde J} M} &\to& \frac{1}{\a} d_{{\tilde I}{\tilde J}
  M} \hfill}\right. \label{scaleinv}
\end{eqnarray}
As we will see in the following, for the $N=2$ theory in five
dimensions this freedom corresponds to the possibility of choosing an
overall normalization for the tensor contributions to the
Chern--Simons Lagrangian.


\section{$D=5$, $N=2$ supergravity revisited}
\label{susy}
\subsection{Generalities and differences from previous approaches}
In this section we are going to apply the general
analysis of section \ref{generalities} to the case of  $N=2$
supergravity theory in five dimensions coupled to vector- and
tensor-multiplets.

The field content of the theory, in the absence of couplings, is
\begin{itemize}
\item the gravity supermultiplet
$$(V_\mu^a, \psi_\mu^A, A^0_\mu)\,,\qquad a =0,1,\dots 4\,, \quad\mu =0,1,\dots 4\,, \quad
  A=1,2$$
  where $V^a_\m$ is the space-time vielbein (with $a$ tangent-space indices and $\mu$ world-indices), $\psi^A_\mu$ the gravitino, with R-symmetry index in the fundamental representation of $Sp(2,\R)$, and $A^0$  the graviphoton;
\item  $n_V$ gauge multiplets
$$(A^i_\mu, \lambda^{iA},\varphi^i)\,,\qquad i=1,\dots n_V\,$$
with $ \varphi^i,\l^{iA}$  the scalar partners of the gauge vectors $A^i$ and the $Sp(2,\R)$-valued gaugini respectively. 
Since the gauge vectors mix in the interacting theory, in the following we will introduce the index $\L =(0,i)=0,1,\dots n_V$  running
over all the gauge-vector indices, that is: $A^\L\equiv (A^0,A^i)$;
\item $n_T$ massless tensor multiplets
$$(B_{M|\mu\nu}, \l^{MA},\varphi^M)\,,\qquad i=1,\dots n_T\,$$
with $ \varphi^M,\l^{MA}$  the scalar and spinor partners respectively
of the tensors $B_M$; 
\item $n_H$ hypermultiplets
$$(q^u, \zeta^{\a})\,, \qquad u=(A\a )=1,\dots 4n_H\,; \quad \a
  =1,\dots 2n_H$$
  where the scalars $q^u$ span a quaternionic manifold of quaternionic
  dimension $n_H$ and their spin-1/2 partners $\zeta^\a$ are labeled
  with an index in the fundamental representation of $Sp(2n_H,\R)$. 
\end{itemize}
Before entering in the explicit construction of the theory, let us
emphasize the differences of our approach with respect to the existing
literature on $D=5$, $N=2$ supergravity. Inspired by the analysis of
the previous section, we are interested in exploiting all the rich
gauge structure underlying the bosonic sector of the model, so we want
to retrieve and possibly to extend the results in the existing
literature by starting with massless tensors and letting them take
mass via the anti-Higgs mechanism. Let us discuss this point in some
more detail than what has already done in the introduction.

While the anti--Higgs mechanism is very well understood at the bosonic
level, to implement it within a supersymmetric theory is a non trivial
task. This is due to the fact that the supersymmetry constraints
require the vectors $A^M$ giving mass to the tensors (in the notations
of section \ref{generalities}) to be related to the tensors themselves
in a non local way, involving Hodge-duality. This relation is codified
in the so-called {\em ``self-duality-in-odd-dimensions"} condition  to
which all the tensor fields in odd-dimensional supergravity theories
have to comply \cite{Townsend:1983xs}: 
\begin{equation}
m^{MN}H_{N|abc} \propto \e_{abcde} F^{M|de}.\label{selfodd}
\end{equation}
In particular, for the five dimensional case the tensors are further required to be complex.

In fact, in the approach currently adopted in the literature \cite{Gunaydin:1983rk,Gunaydin:1983bi,Gunaydin:1984pf,Sierra:1985ax,Lukas:1998yy,Gunaydin:1999zx,Ceresole:2000jd,Bergshoeff:2004kh,Gunaydin:2005bf}, the tensors $B_M$
in the tensor multiplets are taken to be massive (and constrained to satisfy \eq{selfodd}) from the very
beginning, without any tensor-gauge freedom.

Naively, to implement the anti-Higgs mechanism at the supersymmetric level one could think of directly supersymmetrizing the FDA \eq{fda}, and try to give mass to the whole tensor multiplets
by coupling them to $n_T$ extra abelian vector multiplets added to the theory:
\begin{equation}
(A^M_\mu, \chi^{M A},\phi^M),\label{false}
\end{equation}
 where the vectors $A^M$ and the
tensors $B_M$ admit the couplings and gauge invariance as in \eq{fda}
and \eq{gaugefin}. If this would be the case, in  the interacting theory the fields in the extra vector multiplets would
couple to the tensor multiplets and one would
end up with $n_T$ {\em long} massive multiplets.
We found,  however, from explicit calculation  that this is not the case, since  supersymmetry transformations never relate the tensors $B_M$ to the spinors $\chi^{MA}$
nor to the scalars $\phi^M$ in \eq{false}. Then the only way compatible with supersymmetry to couple  $N=2$
supergravity with $n_T$ massive tensors  involves {\em short BPS}
tensor multiplets
$$(B_{M|\mu\nu}, \lambda^{MA},\varphi^M)$$
where the massive tensors $B_M$ (that are complex because of CPT
invariance of the BPS multiplet) have to satisfy \eq{selfodd} (see
eq. \eq{selfmass}).
This is evident for the models having a six dimensional uplift, as
discussed in the introduction, since  for these cases the mass of the
tensors is the  BPS central charge  gauged by the graviphoton $g_{\mu
  5}$.
 Then, in order to understand the $N=2$ supergravity
theory in five dimensions coupled to tensor and vector multiplets as a supersymmetrization of
the FDA discussed in section \ref{generalities}, we will adopt the following strategy: we
start from the massless theory with field content as outlined at the beginning of this
section, but we also introduce $n_T$ extra auxiliary abelian vectors $A^M$ coupled to the
system. The closure of the supersymmetry algebra will then fix their field-strengths,
on-shell, to be the Hodge-dual of the field-strengths of the tensors $B_M$. When the theory
also includes non-abelian gauge multiplets gauging some algebra $G_0$, and the tensor
multiplets are charged under some  representation of $G_0$, then the spin-one part $(B_M,
A^M,A^\L)$ of the bosonic sector is coupled as in \eq{fda}. In this case the closure of the
supersymmetry algebra also involves the non abelian field-strengths and give the set of
constraints \eq{bi1} - \eq{bilast} below.

According to the discussion in section \ref{generalities}, to simplify the notation we will generally use  the index ${\tilde I}
= (\L,M)$, valued in a representation of a group $G\supset G_0$ in the notations of section \ref{generalities}, that runs over all the vectors (including the auxiliary
ones) $$A^{{\tilde I}}\equiv (A^\L,A^M)$$ and over  the scalar sections
$$X^{{\tilde I}}(\varphi^x)\equiv (X^\L,X^M)$$
($G$-valued functions of the scalar fields $\varphi^x \equiv (\varphi^i,\varphi^M)$) which appear in the supersymmetry transformations of the vector and tensor fields.
The world-index $x=1,\dots ,n_V +n_T$ will collectively enumerate the
scalar fields $\varphi^x$ and the spinors $\lambda^{x A}$ both in the
tensor and vector multiplets.

\bigskip
With respect to the analysis of
\cite{Gunaydin:1999zx},
our discussion will be a bit more general as we will include a
non-zero Chern--Simons coupling $d_{\L\S M}$, as in \cite{Bergshoeff:2004kh}.


Another important point, not considered so far in this general
context, concerns the couplings $m^{MN}$. The closure of the FDA \eq{fda}
demands (recalling \eq{closure}) the generators $T_{\tilde{I}
  M}{}^N$ to be related to the couplings $d_{\tilde{I}\tilde{J} M}$
and to the structure constants $f_{\tilde{J}\tilde{K}}{}^{\tilde{I}}$
respectively by
\begin{equation}
T_{\tilde{I} N}{}^M =  m^{\tilde{J}M} \, d_{\tilde{I}\tilde{J}N},
\qquad m^{\tilde{I}N} T_{\tilde{J} N}{}^M =
f_{\tilde{J}\tilde{K}}{}^{\tilde{I}} m^{\tilde{K}M}.
\end{equation}
Setting in the second relation  $\tilde{I}=P$ and $\tilde{J}=\L$,
it is immediate to obtain the following:
\begin{equation}
d_{\L MN} m^{MQ}\, m^{NP} = - d_{\L NM} \, m^{PN}m^{MQ}. \label{cs}
\end{equation}
Eq. \eq{cs} in principle admits two different solutions: either $d_{\L
 MN}$ is symmetric and $m^{MN}$ antisymmetric or the opposite.
But since these couplings enter the Lagrangian of five dimensional
supergravity respectively in the kinetic term for the tensors $m^{MN}
B_M \de B_N$ (which for $m^{MN}$ symmetric is a total derivative)  and in the Chern--Simons term $d_{\tilde I MN} A^{\tilde I} F^M F^N$ (which is zero if $d_{\tilde I MN}$ is antisymmetric in $M$ and $N$),
we are forced to consider only the former solution\footnote{This is
  not necessarily true for other cases, like the four dimensional
  theories, where the equation (\ref{cs}) also seems to allow the
  alternative solution
  \begin{equation}
    d_{{\tilde I} MN}= - d_{{\tilde I} NM}; \quad m^{MN} = m^{NM}.
  \end{equation}}.
Furthermore it should be noted that this same choice
forbids the presence of $d_{MNP}$ couplings, due to eq. \eq{msymm}.

We want to stress, however, that this constraint leaves the freedom for
the tensor mass-matrix $m=-m^T$ to have $n_T$ different eigenvalues
$\pm \i m_\ell$ ($\ell=1,\dots n_T/2$)\footnote{For $n_T$ odd there
  is one extra zero-eigenvalue.
  However,  this case is excluded when
  the theory is embedded in $N=2$ supergravity, since in this case, as
  we already anticipated, the closure of the superalgebra requires a
  self-duality condition \cite{Townsend:1983xs} which needs an even
  number of tensors.}.
As anticipated in the introduction, this is the case, for  example, of the five dimensional theory
obtained by Scherk--Schwarz generalized dimensional reduction
\cite{Scherk:1979zr,Cremmer:1979uq} from the $(2,0)$ theory in six
dimensions \cite{Andrianopoli:2004xu}.  In this theory the mass matrix $m^{MN} $ is in fact
the S-S phase, in the Cartan subalgebra of the global symmetry $SO(n_T)
\subset SO(1,n_T)$, which is the isometry group of the scalar sector
of the tensor multiplets in the $D=6$ parent theory. The five
dimensional theory one obtains in this way is a gauged theory with
flat group given by the semidirect product $U(1) \ltimes R_V$, where
$R_V$ is an $n_V$-dimensional representation of $\rSO(1,n_T)$, and the
$U(1)$ group is gauged by the vector coming from the metric in six
dimensions. As remarked in \cite{Andrianopoli:2004xu}, such a
situation was not considered in previous classifications.

On the other hand, if we take all the eigenvalues of the matrix
$m^{MN}$  equal, which is the case generally considered in the
literature
\cite{Gunaydin:1983rk,Gunaydin:1983bi,Gunaydin:1984pf,Sierra:1985ax,Lukas:1998yy,Gunaydin:1999zx,Ceresole:2000jd,Bergshoeff:2004kh,Gunaydin:2005bf},
then  $m^{MN}$ may be set in the form $m^{MN}=m\, \Omega^{MN}$ where
$m$ is one constant real parameter and $\Omega$  the symplectic
metric. In this case the constraints \eq{cs}  require the generators $T_{\L
I}^{\phantom{ \L I}J}$ to belong to a symplectic representation of the
gauge group $T_\L^T \cdot \O + \O \cdot T_\L =0$.


\subsection{The construction of the theory}

Since our approach involves some generalizations with respect to those
in the existing literature, as discussed above, we have rederived
right from the beginning the theory in its full generality. We have
used the superspace geometric approach as far as the solution of
Bianchi identities is concerned (from which the supersymmetry
transformation laws of the fields follow) and  the superspace
rheonomic Lagrangian for the derivation of the Lagrangian on
space-time. We also tried to use, as much as possible, the notations
existing in the previous literature; however in some cases we found
useful to adopt different normalizations for the fields and couplings
with respect to the seminal papers on the subject
\cite{Gunaydin:1983bi,Gunaydin:1999zx,Ceresole:2000jd}.
A dictionary between our normalizations and those adopted in the
previous papers is given in appendix \ref{rosetta}.

\bigskip

Our starting point, for the construction of the theory, is the
generalization of the bosonic FDA of section \ref{generalities} to a
super-FDA  in superspace. Consequently, we introduce the supergravity
one-forms $\Omega^a{}_b$, $V^a$ and $\Psi^A$ denoting respectively
the spin-connection, the vielbein and the gravitino in superspace
($V^a$ and  $\Psi^A$ spanning a basis on
superspace),  together with their ``supercurvatures'' two-forms
$\cR^a{}_b$, $\mathcal{T}^a$ and $\rho^A$. We further introduce
 zero-forms for the scalars $\varphi^x, q^u$ and spin 1/2 fields
 $\lambda^{xA},\zeta^\alpha$ and their curvatures (covariant
 derivatives) .

The $D=5$, $N=2$ super-FDA is:
\begin{eqnarray}
\cR^a{}_b &=& \de \O^a{}_b - \O^a{}_c \wedge \O^c{}_b
\label{eq:curvature} \\
\mathcal{T}^a &=& \de V^a - \O^a{}_b V^b - \imez \ol \Psi_A \G^a
\Psi^A \label{eq:tors} \\
F^{\tilde I} &=& \de A^{\tilde I} + \mez f_{{\tilde J}{\tilde
    K}}{}^{\tilde I} A^{\tilde J} \wedge A^{\tilde K} + m^{{\tilde
    I}M} B_M + \i X^{\tilde I} \ol \Psi_A \Psi^A \label{eq:allA} \\
H_M &=& \de B_M + T_{\tilde{I} M}{}^N A^{\tilde{I}} \wedge B_N +
d_{{\tilde I}{\tilde J}M} \sx( F^{\tilde I} - \i X^{\tilde I} \ol
\Psi_A \Psi^A \dx) A^{\tilde J} + \nn \\
&& + \i X_M \ol \Psi_A \G_a \Psi^A V^a \label{eq:HI} \\
D \varphi^x &=& \de \varphi^x + k^x_{\tilde{I}} A^{\tilde{I}}
\label{eq:xi} \\
D q^u &=& \de q^u + k^u_{\tilde{I}} A^{\tilde{I}} \label{eq:q} \\
\rho^A &=& \de \Psi^A - \qu \O_{ab} \G^{ab} \Psi^A +\tilde
\omega^A{}_B \Psi^B \label{eq:rho} \\
\na \l^{{x} A} &=& \de \l^{x A} - \qu \O_{ab} \G^{ab} \l^{x A}  +
\tilde \G^x{}_{y} \l^{y A} + \tilde\omega^{A}{}_B \l^{xB}
\label{eq:alllambda} \\
\na \zeta^{\a} &=& \de \zeta^{\a} - \qu \O_{ab} \G^{ab}\zeta^{\a} +
\tilde\D^{\a}{}_{\b} \zeta^{\b}. \label{eq:zeta}
\end{eqnarray}
 In \eq{eq:rho} - \eq{eq:zeta} the gauged connections on the scalar
$\sigma$-models $\cM(\varphi)$ and $\cM_H(q)$ appear,  where $\cM(\varphi)$ is parametrized by the scalars in the vector and tensor multiplets while $\cM_H(q)$ is parametrized by the scalars of the hypermultiplets (the quaternionic sector is unaffected by the presence of tensor multiplets).
They are defined as:
\begin{eqnarray}
\matrix{\tilde\G^x{}_y &=&\G^x{}_y + A^{\tilde I} \partial_y
  k^x_{\tilde I} \qquad\qquad \mbox{ $\sigma$-model connection for
    gauge sector} \mm
\tilde\omega^{AB} &=&\omega^{AB}+ \frac{3}{2} A^{\tilde I}
  \mathcal{P}_{\tilde I}{}^{AB}\quad\qquad \mbox{ $SU(2)$ connection}
  \mm
\tilde\D^{\a}{}_{\b} &=& \D^{\a}{}_{\b} + A^{\tilde I} \partial_v
  k^u_{\tilde I} \, \U_u{}^{{\a}A}\U^{v}{}_{{\b}A}\qquad \mbox{
  $Sp(2n_H)$ connection}.\hfill}\label{gaugedspin}
\end{eqnarray}
Here $\G^x{}_y(\varphi)$ is the Christoffel connection one-form of $\cM(\varphi)$, while $\omega^{AB}(q)$ and $\Delta^\alpha{}_\beta(q)$ are respectively the $Sp(2,\R)$ R-symmetry and the $Sp(2n_H,\R)$ connections on $\cM_H(q)$. Furthermore
$k^x_\L(\varphi)$ and $k^u_\Lambda(q)$ denote the Killing vectors on the two $\sigma$-models. In eq. \eq{gaugedspin} also appear the geometric quantities $\U_u{}^{A\alpha}$  and $ \mathcal{P}_{\tilde I}{}^{AB}$. They are the vielbein ($\U^{A\a} \equiv \U^{A\a}_u dq^u$) and prepotential on $\cM_H$. For their definition and geometric properties, we refer the reader to the standard literature, in particular \cite{Andrianopoli:1996cm,D'Auria:2001kv} where the same notations are used.

 We adopted the following conventions
for raising and lowering  $Sp(2,\R)$ and $Sp(2n_H)$ indices:
\begin{equation}
\xi_A = \epsilon_{AB}\xi^B \,;\quad \xi^A =
-\epsilon^{AB}\xi_B\,;\qquad \xi_{\a} = \IC_{{\a}{\b}}\xi^{\b}
\,;\quad \xi^{\a} = - \IC^{{\a}{\b}}\xi_{\b}
\end{equation}
while the flat space-time indices $a,b$ are raised or lowered with the metric
\begin{equation}
\eta_{ab} = {\mathrm{diag(+,-,-,-,-)}}.
\end{equation}
With these definitions, the explicit construction proceeds by first solving the super-Bianchi's following from  \eq{eq:curvature} - \eq{eq:zeta}:
\begin{eqnarray}
\cR^a{}_b V^b &=&  \i \ol \Psi_A \G^a \rho^A \label{eq:bitors} \\
D \cR^a{}_b &=& 0 \label{eq:curv} \\
D F^{\tilde I} &=& m^{{\tilde I}M} \sx( H_M - \i X_M \ol \Psi_A \G_a
\Psi^A V^a \dx) + \i D X^{\tilde I} \ol \Psi_A \Psi^A - 2 \i X^{\tilde
  I} \ol \Psi_A \rho^A \label{eq:biF} \\
D H_M &=& d_{{\tilde I}{\tilde J}M} \sx( F^{\tilde I} - \i X^{\tilde
  I} \ol \Psi_A \Psi^A \dx) \wedge \sx( F^{\tilde J} - \i X^{\tilde J}
\ol \Psi_B \Psi^B \dx) + \nn \\
&& + \i D X_M \ol \Psi_A \G_a \Psi^A V^a - 2 \i X_M \ol \Psi_A \G_a
\r^A V^a - \frac 12 X_M\ol \Psi_A \G_a \Psi^A\ol \Psi_B \G^a \Psi^B
\label{eq:biHI} \\
D^2 \varphi^x &=& k^x_{\tilde I} \sx( F^{\tilde I} - \i X^{\tilde I}
\ol \Psi_A \Psi^A \dx) \label{eq:bix} \\
D^2 q^u &=& k^u_{\tilde I} \sx( F^{\tilde I} - \i X^{\tilde I} \ol
\Psi_A \Psi^A \dx) \label{eq:biq} \\
\na \rho^A &=& - \qu \cR_{ab} \G^{ab} \Psi^A -\tilde \cR^A{}_B \Psi^B
\label{eq:birho} \\
\na^2 \l^{x A} &=& - \qu \cR_{ab} \G^{ab} \l^{xA} - \tilde\cR^{x}{}_{y}
\l^{yA} - \cR^A{}_B \l^{xB} \label{eq:bialllambda}\\
\na^2 \zeta^{\a} &=& - \qu \cR_{ab} \G^{ab} \zeta^{\a}  -\tilde
\cR^{\a}{}_{\b} \zeta^{\b}  \label{eq:bizeta},
\end{eqnarray}
 where we have defined:
\begin{eqnarray}
D F^{\tilde I} &\equiv & \de F^{\tilde I} + \hat f_{{\tilde J}{\tilde
    K}}{}^{\tilde I} A^{\tilde J} F^{\tilde K} \\
D X^{\tilde I} &\equiv & \de X^{\tilde I} + \hat f_{{\tilde J}{\tilde
    K}}{}^{\tilde I} A^{\tilde J} X^{\tilde K} \\
D H_M &\equiv & \de H_M + \hat T_{{\tilde I} M}{}^N A^{\tilde I} H_N\\
D X_M &\equiv & \de X_M + \hat T_{{\tilde I} M}{}^N A^{\tilde I} X_N
\end{eqnarray}
with the $ \hat f_{{\tilde J}{\tilde K}}{}^{\tilde I}$, $\hat
T_{{\tilde I} M}{}^N $ introduced  in \eq{generalgenerat}, and
\begin{eqnarray}
\tilde\cR^x{}_y &\equiv & \de \tilde\G^x{}_y + \tilde\G^x{}_z
\tilde\G^z{}_y \\
\tilde\cR^A{}_B &\equiv & \de \tilde\omega^A{}_B + \tilde\omega^A{}_C
\tilde\omega^C{}_B \\
\tilde\cR^{\a}{}_{\b} &\equiv & \de \tilde\D^{\a}{}_{\b} +
\tilde\D^{\a}{}_{\g} \tilde\D^{\g}{}_{\b}.
\end{eqnarray}
Eq.s \eq{eq:bitors} - \eq{eq:bizeta} are solved by  parametrizating the supercurvatures on superspace (from which the supersymmetry transformation laws follow) as:
\begin{eqnarray}
\mathcal{T}^a &=& 0 \label{par:tors} \\
\cR_{ab} &=& \hat\cR_{abcd} V^c V^d - \i \ol \Psi_A \G_{[a} \r^A_{b]c} V^c - \frac{\i}{8}
X_{\tilde I} \hat F^{{\tilde I} |cd} \e_{abcde} \ol \Psi_A \G^e \Psi^A -
\imez X_{\tilde I}
\hat F^{\tilde I}_{ab} \ol \Psi_A \Psi^A + \nn \\
&&+ \frac 14 g_{xy}\left( \ol \l^{xA} \G^c \l^{yB} \ol \Psi_A \G_{abc}
\Psi_B -  \ol \l^{xA} \G_{ab} \l^{yB} \ol \Psi_A \Psi_B  - \mez  \ol
\l^{xA} \G_{abc} \l^{yB} \ol \Psi_A \G^c \Psi_B\right) + \nn \\
&& + \i S^{AB} \ol \Psi_A \G^{ab} \Psi_B - \frac{1}{4} \ol \zeta_\a
\G_{abc} \zeta^\a \ol \Psi_A \G^c \Psi^A \label{par:r}\\
F^{{\tilde I}} &=&\hat F^{\tilde I}_{ab} V^a V^b - 2 f^{{\tilde
    I}}_x\ol\Psi_A \G_a \l^{x A} V^a
\label{par:anyA} \\
H_M &=& \hat H_{M|abc} V^a V^b V^c - h_{Mx} \ol \Psi_A \G_{ab} \l^{x A} V^a
V^b \label{par:HI} \\
D \varphi^{x} &=& \hat D_a\varphi^{x} V^a + \ol \Psi_A \l^{x A}
\label{par:anyX} \\
Dq^u &=&  \hat D_a q^u V^a + \U^u{}_{A{\a}} \ol \Psi^A \zeta^{\a} \\
\rho^A &=& \rho^A_{ab} V^a V^b - \frac{1}{8} X_{{\tilde I}} \hat F^{{\tilde
    I} |bc} \sx( \G_{abc} - 4 \eta_{a[b} \G_{c]} \dx) \Psi^A V^a +
S^{AB} \G_a \Psi_B V^a + \nn \\
&&\hskip -2mm + \frac\i 4 g_{xy} \Bigl[ \ol \l^{xA} \G^{b} \l^{yB}
  \sx(\G_{bc} + 2 \eta_{bc} \dx) \Psi_B + \qu \ol \l^{xA} \G^{ab}
  \l^{yB} \sx( \G_{abc} + 4 \G_a \eta_{bc} \dx) \Psi_B \Bigr] V^c +
\nn\\
&& - \frac{\i}{8} \ol \zeta_\a \G_{abc} \zeta^\a \G^{ab} \Psi^A V^c
\label{par:rho} \\
\na \l^{xA} &=& \hat\na_a \l^{xA} V^a + \imez D_a \varphi^x \G^a \Psi^A +
\frac{\i}{4} g^x_{{\tilde I}} \hat F^{{\tilde I}}_{ab} \G^{ab} \Psi^A + \i
W^{xAB} \Psi_B + \nn \\
&& + \frac 14 T^x{}_{yz}\left(-3 \ol \l^{yA} \l^{zB} \Psi_B +  \ol
\l^{yA} \G_a \l^{zB} \G^a \Psi_B + \mez \ol \l^{yA} \G^{ab}
\l^{zB} \G_{ab} \Psi_B\right) \label{par:anylambda} \\
\na \zeta^\a &=&\hat \na_a \zeta^\a V^a + \i  \U_{uA}{}^\a D_a q^u \G^a
\Psi^A + \i \N^\a_A \Psi^A ,\label{par:zeta}
\end{eqnarray}
in terms of a set of scalar-dependent quantities:
$$f^{\tilde I}_x\,,\quad h_{Mx}\,,\quad g_{xy}\,,\quad g^x_{\tilde I}\,,\quad T^x{}_{yz}$$
and  of the fermion-shifts due to the gauging $S^{AB}$, $ W^{xAB} $, $\N^\a_A $. The `hat' on the field-strengths and covariant derivatives denotes the
supercovariant part.
  Eq.s   \eq{eq:bitors} - \eq{eq:bizeta}  give a set of constraints among the quanitites appearing in the parametrizations \eq{par:r} - \eq{par:zeta}. Part of them are reported  below:
  \begin{eqnarray}
f^{\tilde I}_{x} &=&  D_x X^{\tilde I}  \\
h_{Mx} &=& - D_x X_M  \\
D_{(y} f^{\tilde I}_{x)} &=& T^{z}{}_{xy} f^{\tilde I}_{z} +  X^{\tilde I} g_{xy}  \label{bi1}\\
T^z{}_{[xy]}&=& 0\label{t[]}\\
X_M &=& - 2 d_{{\tilde I}{\tilde J}M} X^{\tilde I} X^{\tilde J} \\
\hat H_{M|abc} &=& -\frac 16 a_{M{\tilde I}}\e_{abcde} \hat F^{{\tilde I}|de}\label{selfmass}\\
X^{\tilde I} X_{\tilde J} + f^{\tilde I}_x g^x_{\tilde J} &=& \d^{\tilde I}_{\tilde J}\\
f^{\tilde I}_x W^{x[AB]} &=& \mez m^{{\tilde I}M} X_M \e^{AB} \\
S^{AB} &=& X^{\tilde I} \mathcal{P}_{\tilde I}{}^{AB} \, ; \qquad S^{[AB]} = 0 \\
2 X_M S^{(AB)} &=& h_{M x} W^{x (AB)} \\
\mathcal{P}_{\tilde I}{}^{AB} m^{{\tilde I}M} &=& k_{\tilde I}^u m^{{\tilde I}M} = k_{\tilde I}^x m^{{\tilde I}M} = 0  \\
\mathcal{P}_{[{\tilde I}}{}^{AC} \mathcal{P}_{{\tilde J}]C}{}^B &=& \frac{1}{3}
f_{{\tilde I}{\tilde J}}{}^{\tilde K} \mathcal{P}_{\tilde K}{}^{AB} \label{bilast} .
\end{eqnarray}
In eq. \eq{selfmass} we have introduced the matrix
\begin{eqnarray}
a_{{\tilde I}{\tilde J}}&\equiv& X_{\tilde I} X_{\tilde J} +
h_{{\tilde I} x} g^x_{\tilde J}\label{aij}
\end{eqnarray}
which will appear in the Lagrangian as kinetic matrix for the vector
field-strengths.
Eq. \eq{selfmass} expresses, at the supersymmetric level, the duality
relation among the $B$-field-strengths and the vector
field-strengths. 

Since the analysis has been done only at 2-fermion level, these are
not the totality of the algebraic and geometric
constraints of the theory. Further constraints are more easily
evaluated from the equations of motion in superspace of the rheonomic
Lagrangian given in appendix \ref{rheonomic} and will be reported
in the next subsection.


\subsection{The Lagrangian}
\label{lagrangian}
Writing the action as:
\begin{equation}
S =\int \sqrt{-g} {\de}^5 x \cL ,
\end{equation}
the Lagrangian of the theory is:
\begin{eqnarray}
\cL &=& \cL_{Grav}+\cL_{Kin} + \cL_{Pauli} + \cL_{gauge} + \cL_{CS}
+\cL_{4f} \label{susylag}
\end{eqnarray}
with:
\begin{eqnarray}
\cL_{Grav} &=& R + \frac{\i}{\sqrt{-g}}\ol \Psi_{A|\m} \G^{\m\n\rho} \rho^A_{\n\rho}
\label{eq:lgrav} \\
\cL_{Kin} &=& - \frac{3}{8} a_{{\tilde I}{\tilde J}}
\mathcal{F}^{\tilde I}_{\m\n} \cF^{{\tilde J}|\m\n} +  \frac3{16}\frac 1{\sqrt{-g}}\epsilon^{\m\n\r\s\l} m^{MN}
B_{M|\m\n}D_\r B_{N|\s\l}+ \nn\\
&&+\frac{3}{4} g_{xy} D_\m \varphi^x D^\m \varphi^y + g_{uv} D_\m q^u
D^\m q^v+ \nn\\
&&+\frac{3}{2} \i g_{xy} \ol \l^{x}_A \G^\m \na_\m \l^{y A} + \i \ol
\zeta_{\a} \G^\m D_\m \zeta^{\a} \label{eq:lkin} \\
\cL_{Pauli} &=&- \frac{3}{8} \i X_{\tilde I} \cF^{\tilde I}_{\m\n} \ol
\Psi_{A|\rho}\G^\m \G^{\rho\sigma} \G^\n \Psi^A_\sigma + \frac{3}{4}
h_{{\tilde I} x} \cF^{\tilde I}_{\m\n} \ol \Psi_{A|\rho}
\left(\G^{\m\n\rho}-2\G^{\n}g^{\m\rho}\right) \l^{xA} + \nn \\
&& + \frac{\i}{4} \phi_{{\tilde I} xy} \cF^{\tilde I}_{\m\n} \ol
\l^x_A \G^{\m\n} \l^{yA}- \frac{3}{2} g_{xy} D_\n \varphi^x \ol
\Psi_{A|\m}\G^\n\G^\m \l^{yA} +\nn\\
&& + \frac{3\i}{4} X_{\tilde{I}} \cF^{\tilde{I}}_{\m\n} \ol \zeta_{\a}
\G^{\m\n} \zeta^{\a} - 2 D_\n q^u \ol \Psi_{A|\m} \G^\n\G^\m
\zeta_{\a} \U_u{}^{A{\a}} \label{eq:lpauli} \\
\cL_{gauge} &=& - 3 \i S^{AB} \ol \Psi_A^\m \G_{\m\n} \Psi_B^\n - 3
g_{xy} W^{y AB} \ol \l_A^x \G_\m \Psi_B^\m +2 \N^{A}_{\a} \ol
\Psi_{A|\m} \G^\m \zeta^{\a} + \nn \\
&& - \frac{3}{2} \i M_{xy|AB} \ol \l^{xA} \l^{yB} - \i \M^{{\a}{\b}}
\ol \zeta_{\a} \zeta_{\b} - 2  \i \M^{A{\a}}_x \ol \zeta_{\a} \l^x_A
- V(\phi) \label{eq:gauge}\\
\cL_{CS} &=& \frac3{16}\Bigl[ 2 m^{MN} B_{M\m\n} d_{{\tilde I}{\tilde J}
    N} \left( F^{\tilde I}_{\rho\sigma} -\i X^{\tilde I}
    \bar\Psi_{A|\rho} \Psi_\sigma^A \right) A_{\tau}^{\tilde J}  + \nn\\
&& + \frac 13 t_{{\tilde I}{\tilde J}{\tilde K}} A^{\tilde I}_\mu \der_\nu
A^{\tilde J}_\rho \der_\sigma A^{\tilde K}_\tau + \frac 1{4}
    \left(t_{{\tilde I}{\tilde L}{\tilde M}} f_{{\tilde J}{\tilde
    K}}{}^{\tilde L} +  4d_{ {\tilde I} {\tilde J} M} m^{MN}
    d_{{\tilde M}  {\tilde K} N}\right) A_\m^{\tilde I} A_\n^{\tilde
    J} A_\rho^{\tilde K} \der_\sigma A_\tau^{\tilde M} + \nn \\
&&+\frac{1}{20}\left( t_{{\tilde I}{\tilde L}{\tilde M}} f_{{\tilde
    J}{\tilde K}}{}^{\tilde L} + 4 d_{{\tilde I}{\tilde J}M} m^{MN}
d_{{\tilde M} {\tilde K} N} \right)f_{{\tilde N}{\tilde P}}{}^{\tilde
  M} A_\m^{\tilde I} A_\n^{\tilde J} A_\rho^{\tilde K}
    A_\sigma^{\tilde N} A_\tau^{\tilde P} \Bigr]
    \frac{\e^{\m\n\rho\sigma\tau}}{\sqrt{-g}} 
\label{csbis}
\end{eqnarray}
where:
\begin{eqnarray}
M_{xy|AB} &=& (g_{yz} k^z_{\tilde I} f^{\tilde I}_x - \mez h_{Mx}
m^{MN} h_{Ny}) \e^{AB} -  2f^{\tilde I}_z T^z{}_{xy}
\mathcal{P}_{\tilde I}{}^{AB}\\
\M^{{\a}{\b}} &=& \mez \U_v{}^{A\a} \U_{uA}{}^{\b} D^{[u}
  k^{v]}_{\tilde{I}} X^{\tilde{I}} \\
\M^{A{\a}}_x &=& -2 \,\U_u{}^{A{\a}} k^u_{\tilde{I}} f^{\tilde{I}}_x
\end{eqnarray}
and:
\begin{eqnarray}
\left\{\matrix{S^{AB} &=& X^{\tilde I} \mathcal{P}_{\tilde I}{}^{AB}
  \hfill\cr
W^{xAB} &=&  g^{xy} (\frac 12 h_{{\tilde I} y} m^{{\tilde I}M} X_M
  \e^{AB} - 2 f^{\tilde I}_y \mathcal{P}_{\tilde I}{}^{AB} )\hfill\cr
\N^{A{\a}} &=& 2 \U_u{}^{A{\a}} k^u_{\tilde{I}} X^{\tilde{I}}
  .\hfill}\right. \,. \label{shifts}
\end{eqnarray}
Finally, $t_{\tilde I\tilde J \tilde K}$ introduced in \eq{csbis} is a covariantly constant tensor.

In \eq{csbis}, the freedom under rescaling \eq{scaleinv} has been used
to fix the overall normalization. More details on the calculation are
given in appendix \ref{rheonomic}. The 4-fermions contributions to the Lagrangian, from  \cite{Gunaydin:1983bi}, is reported in appendix \ref{4f}.

The scalar potential is
\begin{eqnarray}
V  &=& -12 S^{AB} S_{AB} + \frac 32 g_{xy} W^{x AB} W^y_{AB} +
\N_{{\a}A}\N^{{\a}A}\nn\\
&=& 6 \mathcal{P}_{\tilde{I}}^{AB}\mathcal{P}_{{\tilde{J}} AB}
\left(f^{\tilde{I}}_x g^{xy}f^{\tilde{J}}_y - 2X^{\tilde{I}}
X^{\tilde{J}}\right) + \frac{3}{4} X_M X_N m^{MP}
m^{NL}h_{P x}g^{xy}h_{Ly} + \nn\\
&&+4 g_{uv}k^u_{\tilde{I}} k^v_{\tilde{J}} X^{\tilde{I}} X^{\tilde{J}}
\label{potential}.
\end{eqnarray}
The following Ward-identity on the gauging holds
\begin{equation}
V \d^B_A = - 24 S^{BC} S_{CA} + 3 g_{xy} W^{x CB} W^y_{CA} +
2\N_{{\a}A}\N^{{\a}B}.\label{ward}
\end{equation}
Eq. \eq{ward} is identically satisfied, given eq.s \eq{shifts}, for
any $SU(2)$-valued $\mathcal{P}^{AB}_{\tilde{I}}
=\mathcal{P}^r_{\tilde{I}} \sigma_r^{AB}$.

The Lagrangian \eq{susylag} is left invariant by the  supersymmetry transformation
rules (with supersymmetry parameter $\e^A$):
\begin{eqnarray}
\matrix{ \d V^a_\mu &=& - \i \ol\Psi_{A\mu}\G^a \e^a \hfill \cr
\d A^{\tilde I}_\mu &=& 2\i X^{\tilde I} \ol\Psi_{A\mu} \e^A - 2
f^{\tilde I}_x \ol\e_A \G_\mu \l^{xA} \hfill \cr
\d B_{M \m\n} &=& 2\i d_{{\tilde I}{\tilde J}M} X^{\tilde I} A^{\tilde
  J}_{[\m}\ol\Psi_{\n] A} \e^A + 2\i X_M \ol\Psi_{A[\m}\G_{\n]}\e^A -
h_{Mx} \ol\e_A \G_{\m\n}\l^{xA} \hfill \cr
\d \varphi^x &=& \ol\e_A \l^{xA}  \hfill \cr
\d q^u &=& \U^u{}_{A{\a}} \ol \e^A \zeta^{\a} \hfill\cr
\d \Psi^A_\m &=& D_\m \Psi^A - \omega^A_{u|\,B}\U^u{}_{C{\a}} \ol \e^C
\zeta^{\a}\Psi^B_\m- \frac{1}{8} X_{{\tilde I}} F^{{\tilde I} |\n\r}
\sx( \G_{\m\n\r} - 4 \eta_{\m[\n} \G_{\r]} \dx) \e^A  + S^{AB} \G_\m \e_B
 + \hfill\cr
&&\hskip -2mm + \frac\i 4 g_{xy} \Bigl[ \ol \l^{xA} \G^{\n} \l^{yB}
   \sx(\G_{\n\m} + 2 \eta_{\n\m} \dx) \e_B + \qu \ol \l^{xA} \G^{\n\r}
   \l^{yB} \sx( \G_{\m\n\r} + 4 \G_\n \eta_{\r\m} \dx) \e_B \Bigr]
 \hfill\cr
&& - \frac{\i}{8} \ol \zeta_\a \G_{\mu\nu\sigma} \zeta^\a
 \G^{\nu\sigma} \epsilon^A \hfill\cr
\d \l^{xA} &=& -\omega^A_{u|\,B}\U^u{}_{C{\a}} \ol \e^C \zeta^{\a}\l^{xB}
- \G^x_{yz}\ol\e_C\l^{zC}\, \l^{yA} + \imez D_\m \phi^x \G^\m \e^A +
\frac{\i}{4} g^x_{{\tilde I}} F^{{\tilde I}}_{\m\n} \G^{\m\n} \e^A +
\i W^{xAB} \e_B + \hfill\cr
&& + \frac 14 T^x{}_{yz}\left(-3 \ol \l^{yA} \l^{zB} \e_B +  \ol \l^{yA}
\G^\m \l^{zB} \G_\m \e_B + \mez \ol \l^{yA} \G^{\m\n}
\l^{zB} \G_{\m\n} \e_B\right) \hfill\cr
\d \zeta^\a &=& -\D^{\a}_{u|\,\b}\U^u{}_{A{\g}} \ol \e^A \zeta^{\g}
\zeta^{\b}+ \i D_\m q^u \G^\m U_{uA}{}^\a \e^A + \i \N^\a_A \e^A\hfill
}\label{susyrules}
\end{eqnarray}

For our calculations, we used the geometrical (rheonomic) approach
which, as is well known, provides not only the space-time Lagrangian
and the superspace equations of motion, but also the value of the
generalized curvatures in superspace thus providing constraints on the
physical fields of the theory. This is of course equivalent to require
space-time supersymmetry.

 The extra constraints we find besides those
already given by closure of the Bianchi identities
\eq{bi1}-\eq{bilast} are:
\begin{eqnarray}
t_{{\tilde I}{\tilde J}{\tilde K}} X^{\tilde I} X^{\tilde J} X^{\tilde
  K} &=&1 \label{surface0}\\
X_{\tilde I} &=&   t_{{\tilde I}{\tilde J}{\tilde K}} X^{\tilde J}
  X^{\tilde K} = a_{{\tilde I}{\tilde J}} X^{\tilde J} \\
f^{\tilde I}_x &=& D_x X^{\tilde I}\\
g_{xy}&=&- 2  t_{{\tilde I}{\tilde J}{\tilde K}}X^{\tilde K} f^{\tilde
  I}_x f^{\tilde J}_y =a_{{\tilde I}{\tilde J}}f^{\tilde I}_x
  f^{\tilde J}_y\\
T^z_{xy} &=&  t_{{\tilde I}{\tilde J}{\tilde K}} g^{zw} f^{\tilde I}_w
  f^{\tilde J}_x f^{\tilde K}_y \\
h_{{\tilde I} x} &=& - D_x X_{\tilde I} = a_{{\tilde I}{\tilde J}}
  f_x^{\tilde J} = g_{xy} g_{\tilde I}^y \\
t_{{\tilde I}{\tilde J}{\tilde K}} X^{\tilde K} &=&-\frac{1}{2}
  (a_{{\tilde I}{\tilde J}}-3X_{\tilde I} X_{\tilde J})\\
X_{\tilde I} X^{\tilde I} &=& 1 \\
D_x h_{{\tilde I} y} &=& -  ( h_{{\tilde I} z} T^z_{xy} +  X_{\tilde
  I} g_{xy}) \,; \qquad
D_{[x} h_{{\tilde I} | y]} = 0 \\
\phi_{{\tilde I} xy} &=& 3 t_{{\tilde I}{\tilde J}{\tilde K}}
  f^{\tilde J}_x f^{\tilde K}_y + \frac 94 X_{\tilde I} g_{xy} \\
d_{({\tilde I}{\tilde J} )M} &=& -\frac 12 t_{{\tilde I}{\tilde J}M}\label{3.54}\\
h_{{\tilde I} z}T^z_{xy} &=&  t_{{\tilde I}{\tilde J}{\tilde K}}
  f^{\tilde J}_x f^{\tilde K}_y + \frac 12 X_{\tilde I} g_{xy}.
\end{eqnarray}
In particular,   \eq{surface0} defines the equation of the surface generally carachterizing the scalar geometry of $D=5$, $N=2$ tensor and vector multiplet sector.
Furthermore, the above relations also imply the constraints on the curvature of $\cM(\varphi)$  characterizing its geometry:
\begin{eqnarray}
R^x_{\ yzt}&=& \left( \delta^x_{[t}g_{z]y}+ T^x_{\ w[t}T^w_{\
    z]y}\right)\label{curva}\end{eqnarray}
 and a relation between the  constant $t_{{\tilde I}{\tilde J}{\tilde K}}$ defining the surface and the scalar-dependent couplings:
 \begin{eqnarray}
t_{{\tilde I}{\tilde J}{\tilde K}}&=&\frac 12 \left(5 X_{\tilde I}
X_{\tilde J} X_{\tilde K} -3 a_{({\tilde I}{\tilde J}} X_{{\tilde K}
  )} + 2 T_{xyz}g^x_{\tilde I} g^y_{\tilde J} g^z_{\tilde K}\right).
\end{eqnarray}
Since the geometrical properties of the $\sigma$-model $\cM(\varphi)$ have been discussed thoroughly in the original paper \cite{Gunaydin:1983bi}, we omit further comments on this point.


\subsection{Comments on the scalar potential}

The scalar potential that we find in eq. \eq{potential}:
\begin{eqnarray}
V  &=&  6 \mathcal{P}_{\tilde{I}}^{AB}\mathcal{P}_{{\tilde{J}} AB}
\left(f^{\tilde{I}}_x g^{xy}f^{\tilde{J}}_y - 2X^{\tilde{I}}
X^{\tilde{J}}\right) + \frac{3}{4} X_M X_N m^{MP} m^{NL}h_{P
  x}g^{xy}h_{Ly} + \nn \\
&&+4 g_{uv}k^u_{\tilde{I}} k^v_{\tilde{J}} X^{\tilde{I}} X^{\tilde{J}}
\label{potential2}
\end{eqnarray}
is formally the same as the one found in the literature
\cite{Ceresole:2000jd} (a rescaling of the fields is required for a
precise comparison;  a map is given in Appendix
\ref{GST/ADS}). However, since we are considering more general
couplings and a non-trivial $m^{MN}$ matrix, a few comments are in
order.

First of all, it is already known that the presence of the tensor multiplets allows a non-zero
$W^{x[AB]}= \frac 12 g^{xy} h_{{\tilde I} y} m^{{\tilde I}M} X_M \e^{AB}$ but that  the contribution to the scalar potential coming from the tensors is always  positive, so that  Anti de Sitter solutions may only be accounted
for a non trivial (possibly constant) $\mathcal{P}_{\tilde{I}}^{AB}$,
giving a mass to the gravitino, while, in the case
$\mathcal{P}_{\tilde{I}}^{AB}=0$, only Minkowski vacua are attainable,
for
\begin{equation}h_{Mx} m^{MN} X_N =0 .\label{vac}
\end{equation}
This is in particular the case when one considers as $N=2$ model
the Scherk--Schwarz generalized dimensional reduction of a six
dimensional theory, as discussed in \cite{Andrianopoli:2004xu}.
For the case
of the S-S dimensionally reduced theory, to have a  non
negative  scalar potential a cancellation is needed between the gaugino and gravitino
contributions, proportional to the prepotential $\mathcal{P}_{\tilde
  I}^{AB}$. This does not
appear instead to be necessary in more general, purely five dimensional,
cases, still allowing, however, a general antisymmetric matrix $m^{MN}$.
Then, when $m^{MN}$ has general skew-eigenvalues, eq. \eq{vac} may have solutions more general than the ``symplectic-orthogonality'' condition between $h_{Mx}$ and $X_N$.

Let us now see the implications of having $d_{\L\S M} \neq
0$. Eq. \eq{vac} has a solution for:
\begin{eqnarray}
X_M  = t_{{\tilde I}{\tilde J}M}X^{\tilde I} X^{\tilde J} = -4 d_{\L
  NM}X^\L X^N -2 d_{\L\S M}X^\L X^\S =0 \label{vacuumtens}
\end{eqnarray}
where we used the relation  \eq{3.54}
Eq. \eq{vacuumtens} must be solved together with the defining equation
of the scalar geometry
\begin{eqnarray}
t_{{\tilde I}{\tilde J}{\tilde K}} X^{\tilde I} X^{\tilde J} X^{\tilde
  K}  = 1
\end{eqnarray}
that is:
\begin{eqnarray}
X^\L \left(t_{\L\S\G} X^\S X^\G +2 t_{\L\S M} X^\S X^M + 2 t_{\L MN}
X^M X^N  \right)= 1 .\label{surface}
\end{eqnarray}
 In \eq{surface} we used the fact that, for $m^{MN}$ invertible, $t_{MNP} =0$, as explicitly shown in appendix \ref{rheonomic}, eq. \eq{tmnp0}. 
Eq. \eq{surface} requires $X^\L \neq 0$ for at least one value of $\L$
(e.g. $X^\L|_{\mbox{vac}} \propto \d^\L_0$), and it then implies that
the v.e.v. of the scalars $X^M$ are now shifted from zero, since
eq. \eq{vacuumtens} is solved for
\begin{equation}
d_{\L MN} X^N|_{\mbox{vac}} =-\frac 12 d_{\L\S M}
X^\S|_{\mbox{vac}}\neq 0.
\end{equation}

\section{Conclusions and outlook}
In the present paper we have studied the $D=5$, $N=2$ theory coupled
to vector, tensor and hyper multiplets by including all possible
couplings compatible with gauge symmetry and supersymmetry. We paid
particular attention in analyzing the algebraic structure of the FDA
which  underlies the theory, This allowed to relax some constraints on the couplings usually considered, and correspondingly to write-down a scalar potential a bit more general than usually considered. It would be interesting to analyze in detail the critical points of models exhibiting the features described here, as in particular a magnetic coupling $m^{MN}$ with arbitrary skew-eigenvalues. Models of this kind (an example of which is found by Scherk--Schwarz compactification from six dimensions \cite{Andrianopoli:2004xu}) should appear in general flux compactifications from superstring or M-theory.

Our investigation may now be extended in various directions.
At a group-theoretical point of view, it would be interesting to extend the FDA to include also higher order forms, as is the case, in general, in theories corresponding to compactifications from superstrings or M-theory.
 We would also like to perform an analysis, on the same lines of the one presented here, for the $D=4$ $N=2$ theory coupled to vector-tensor
multiplets. These developments are under investigation and will be
discussed elsewhere.

\acknowledgments{It is a pleasure to acknowledge valuable discussions with Mario Trigiante all over the preparation of this paper, and  useful observations and careful reading from Maria Antonia Lled\'o on the group theoretical part.
 Work supported in part by the European Community's Human Potential
Program under contract MRTN-CT-2004-005104 `Constituents,
fundamental forces and symmetries of the universe', in which L.A., and
R.D'A.  are associated to Torino University. Partial support from the Spanish Ministry of
  Education and Science (FIS2005-02761) and EU FEDER funds}


\appendix

\section{A trivial deformation of the FDA}\label{deformedfda}
In this appendix we show that a further possible deformation of the FDA \eq{fda} via an extra
3-vector contribution in the field strengths $H_M$ can be always reabsorbed by a field redefinition provided we do not couple the system to higher order forms.
It is in fact  possible to 
deform the FDA \eq{fda} as follows
\begin{eqnarray}
\left\{
\matrix{F^{\tilde I} &\equiv& \de A^{\tilde I} + \mez f_{{\tilde
      J}{\tilde K}}{}^{\tilde I} A^{\tilde J} \wedge A^{\tilde K} +
  m^{{\tilde I} M} B_M \mm
H_M &\equiv& \de B_M + T_{{\tilde I} M}{}^N A^{\tilde I} B_N +
d_{{\tilde I}{\tilde J} M} F^{\tilde I} \wedge A^{\tilde J}+
e_{M{\tilde I}{\tilde J}{\tilde K}} A^{\tilde I} \wedge A^{\tilde J}
\wedge A^{\tilde K}\hfill} \right. \label{fdadeformed}
\end{eqnarray}
with the constant $e_{M{\tilde I}{\tilde J}{\tilde K}}=e_{M[{\tilde
      I}{\tilde J}{\tilde K}]} $ completely antisymmetric in the last
3 indices. This is a  deformation of the FDA structure, which leaves
unchanged the Bianchi identities \eq{BI}, but modifies the constraints in the following way:
\begin{eqnarray}
\matrix{f_{[{\tilde I}{\tilde J}}{}^{\tilde L}  f_{{\tilde K}]{\tilde
      L} }{}^{\tilde M} &=& 2 e_{M{\tilde I}{\tilde J}{\tilde
      K}}m^{{\tilde M}  M} \mm
\sx[ T_{\tilde I} , T_{\tilde J} \dx]_M{}^P &=& f_{{\tilde I}{\tilde
      J}}{}^{\tilde K} T_{{\tilde K} M}{}^P + 6 e_{M{\tilde I}{\tilde
      J}{\tilde K}}m^{{\tilde K} P}\mm
T_{{\tilde I} M}{}^{(N} m^{{\tilde I} |P)} &=& 0 \mm
m^{{\tilde I} N} T_{{\tilde J} N}{}^M &=& f_{{\tilde J}{\tilde
      K}}{}^{\tilde I} m^{{\tilde K} M} \mm
T_{{\tilde I} M}{}^N &=& d_{{\tilde I}{\tilde J} M} m^{{\tilde J} N}
      \mm
T_{[{\tilde I} | M}{}^N d_{{\tilde K} | {\tilde J}] N} &-& \hat f_{ [
      {\tilde I} | {\tilde K}}{}^{\tilde L}   d_{{\tilde L}  | {\tilde
      J}] M} - \mez f_{{\tilde I}{\tilde J}}{}^{\tilde L}  d_{{\tilde
      K}{\tilde L}  M} + 3e_{M{\tilde I}{\tilde J}{\tilde K}} = 0 \mm
e_{N[{\tilde J}{\tilde K}{\tilde L}} T_{{\tilde I}]M}{}^N &-&
      \frac{3}{2} e_{M{\tilde P} [{\tilde K} {\tilde L} }f_{{\tilde
      I}{\tilde J}]}{}^{\tilde P} =0.\hfill} \label{closuredeformed}
\end{eqnarray}
The last equation of \eq{closuredeformed} means that $e_{M {\tilde
    I}{\tilde J}{\tilde K}}$ is a cocycle of the Lie algebra $G$.

 For non-zero $e_{M{\tilde I}{\tilde J}{\tilde K}}$, the gauge
 transformations of the system are deformed into
\begin{eqnarray}
\left\{\matrix{\delta A^{\tilde I} &=& \de \e^{\tilde I} + f_{{\tilde
      J}{\tilde K}}{}^{\tilde I} A^{\tilde J} \e^{\tilde K} -
  m^{{\tilde I} M} \L_M \mm
\delta B_M &=& \de \L_M + T_{{\tilde I} M}{}^N A^{\tilde I} \L_N -
      d_{{\tilde I}{\tilde J} M} F^{\tilde I} \e^{\tilde J} -
T_{{\tilde I} M}{}^N \e^{\tilde I} B_N - 3 e_{M{\tilde I}{\tilde
      J}{\tilde K}} A^{\tilde I} \wedge A^{\tilde J} \e^{\tilde
      K}\hfill}\right.
\end{eqnarray}
but give, for the  field strengths, the same gauge transformation of
the undeformed theory:
\begin{eqnarray}
\left\{\matrix{\delta F^{\tilde I} &=& - \hat f_{{\tilde J}{\tilde
      K}}{}^{\tilde I}  \e^{\tilde J} F^{\tilde K} \mm
\delta H_M &=& - \hat T_{{\tilde I} M}{}^N  \e^{\tilde I} H_N. \hfill}
      \right. \label{labfalfa2}
\end{eqnarray}
 As we see from
\eq{closuredeformed}, in this case the Jacobi identities fail to
close and the $T_{{\tilde I} M}{}^N$ do not generate anymore the algebra
$G$, which is explicitly broken.  However,  for any
general value of $e_{M{\tilde I}{\tilde J}{\tilde K}}$ subject to
\eq{closuredeformed}, the entire algebra $G$ is still generated by the
hatted generators $\hat f$, $\hat
T$, that still satisfy eq.s \eq{bigalgebra}. The consistency of the extended
theory is guaranteed since, from \eq{bigalgebra} we have, for any
$e_{M{\tilde I}{\tilde J}{\tilde K}}$:
\begin{eqnarray}
\matrix{ \left[ \left[\hat f_{[{\tilde I}},\hat f_{\tilde
      J}\right],\hat f_{{\tilde K} ]}\right]_{\tilde L}{}^{\tilde
    P}&=&-f_{[{\tilde I}{\tilde J}}{}^{\tilde N} f_{{\tilde K}]{\tilde
      N} }{}^{\tilde M}  \hat f_{{\tilde M} {\tilde L} }{}^{\tilde P}
  =-2 e_{M{\tilde I}{\tilde J}{\tilde K}} m^{{\tilde N}  M} \hat
  f_{{\tilde N} {\tilde L} }{}^{\tilde P} =0 \mm
 \left[ \left[\hat T_{[{\tilde I}},\hat T_{\tilde J}\right],\hat
      T_{{\tilde K} ]}\right]_M{}^N&=&-f_{[{\tilde I}{\tilde
      J}}{}^{\tilde M} f_{{\tilde K}]{\tilde M}}{}^{\tilde L} \hat
      T_{{\tilde L}M}{}^N =-2 e_{P{\tilde I}{\tilde J}{\tilde K}}
      m^{{\tilde L} P} \hat T_{{\tilde L} M}{}^N =0\hfill} \label{e}
\end{eqnarray}
due to \eq{closuredeformed} and in particular to
\begin{eqnarray}
\matrix{ m^{{\tilde I} N} T_{{\tilde J} N}{}^M -  f_{{\tilde J}{\tilde
 K}}{}^{\tilde I} m^{{\tilde K} M}=0&\to & \hat f_{{\tilde K}{\tilde
 J}}{}^{\tilde I} m^{{\tilde K} M}=0 \mm
T_{{\tilde I} M}{}^{(N} m^{{\tilde I} |P)} = 0&\to & \hat T_{{\tilde
 I} M}{}^{N} m^{{\tilde I} P} = 0.\hfill}
\end{eqnarray}
Therefore we can state that the gauge algebra $G$, even if not
anymore realized in an abstract way, still closes when acting on the
physical generators $\hat f$, $\hat T$  appearing in the Bianchi
identities. To complete the proof that the extension of the FDA
\eq{fda} to include the $e_{M\tilde I \tilde J \tilde K}$ is trivial,
we are now going to show that, when  expressed only in terms of the
physical couplings, the structure of the FDA is not affected by any
possible contribution in $e_{M\tilde I \tilde J \tilde K}$.
To do so, let us recast the theory in terms of the physical couplings
appearing in the Bianchi identities \eq{BI}, as we did in section
\ref{fdageneralities} for the FDA \eq{fda}.  As shown in section
\ref{fdageneralities}, this is done
by the field redefinition \ref{bridef}. Then the FDA \eq{fda3} takes the form:
\begin{eqnarray}
\left\{\matrix{F^{\tilde I} &\equiv& \de A^{\tilde I} + \mez \hat
  f_{{\tilde J}{\tilde K}}{}^{\tilde I} A^{\tilde J} \wedge A^{\tilde
    K} + m^{{\tilde I} M}\tilde B_M \mm
H_M &\equiv& \de \tilde B_M + \frac 12 \hat T_{{\tilde I} M}{}^N
  A^{\tilde I} \tilde B_N + d_{({\tilde I}{\tilde J}) M} F^{\tilde I}
  \wedge A^{\tilde J} + \mathcal{K}_{M {\tilde I}{\tilde J}{\tilde
  K}}A^{\tilde I}\wedge A^{\tilde J} \wedge A^{\tilde
  K}\hfill}\right. \label{fda3}
\end{eqnarray}
where:
\begin{equation}
\mathcal{K}_{M{\tilde I}{\tilde J}{\tilde K}}= e_{M{\tilde I}{\tilde
    J}{\tilde K}} -\frac 12  T_{[{\tilde I} M}{}^Nd_{{\tilde J}{\tilde
      K}]N} - \frac 14 \hat f_{[{\tilde I}{\tilde J}}{}^{\tilde{L}}
  d_{{\tilde K}] {\tilde{L}} M} + \frac 14 d_{{\tilde{L}} [{\tilde K}
    M} \hat f_{{\tilde I}{\tilde J} ]}{}^{\tilde{L}}, \label{ke}
\end{equation}
and:
\begin{eqnarray}
\matrix{\tilde f_{[{\tilde I}{\tilde J}}{}^{\tilde M} \tilde
    f_{{\tilde K}]{\tilde M}}{}^{\tilde L} &=& 2 m^{{\tilde L} M}
  \mathcal{K}_{M[{\tilde I}{\tilde J}{\tilde K} ]} \mm
\hat T_{[{\tilde I} M}{}^N  \hat T_{{\tilde J} ]N}{}^P  &=& \tilde
    f_{{\tilde I}{\tilde J}}{}^{\tilde K} \hat T_{\tilde K} + 12
    \mathcal{K}_{M{\tilde I}{\tilde J}{\tilde K}}m^{{\tilde K} P} \mm
\hat T_{{\tilde I} M}{}^{N} m^{{\tilde I} P} &=& 0\mm
\frac{1}{2} m^{{\tilde I} N} \hat T_{{\tilde J} N}{}^M &=&\tilde
    f_{{\tilde J}{\tilde K}}{}^{\tilde I} m^{{\tilde K} M} \mm
\hat T_{{\tilde I} M}{}^N &=& 2 d_{({\tilde I}{\tilde J}) M}
    m^{{\tilde J} N} \mm
\hat T_{[{\tilde I} | M}{}^N d_{( {\tilde J}]{\tilde K} )N} &-& 2\hat
    f_{ [ {\tilde I} | {\tilde K}}{}^{\tilde L} d_{ ({\tilde
    J}]{\tilde L}) M} -\tilde f_{{\tilde I}{\tilde J}}{}^{\tilde L}
    d_{({\tilde K}{\tilde L}) M} = -6 \mathcal{K}_{M{\tilde I}{\tilde
    J}{\tilde K}} \mm
\mathcal{K}_{N [{\tilde J}{\tilde K}{\tilde L}} \hat T_{{\tilde
    I}]|M}{}^N&-&3  \mathcal{K}_{M {\tilde P} [{\tilde I}{\tilde
    J}}\tilde f_{{\tilde K}{\tilde L}]}{}^{\tilde P} =0.\hfill}
\label{closure3}
\end{eqnarray}
Note that, by substituting the value of $e_{M{\tilde I}{\tilde J}{\tilde K}}$
given by \eq{closuredeformed}, eq. \eq{ke} may be rewritten as
\begin{equation}
\mathcal{K}_{M{\tilde I}{\tilde J}{\tilde K}}= \frac 12 \hat
f_{[{\tilde I}{\tilde J}}{}^{\tilde{L}} d_{({\tilde K}] {\tilde{L}})
  M} +\frac{2}{3} d_{({\tilde{L}}[{\tilde J})M} \hat f_{{\tilde
      I}]{\tilde K}}{}^{\tilde{L}}, \label{k2}
\end{equation}
which is identical to \eq{k}. This shows that, when the FDA is expressed
only in terms of the physical couplings $\hat f_{\tilde I}$, $\hat
T_{\tilde I}$ and $d_{({\tilde I}{\tilde J} )M}$, it does not depend on
any possible contribution from $e_{M{\tilde I}{\tilde J}{\tilde K}}$
in \eq{fdadeformed}.  We conclude that the couplings $e_{M{\tilde
    I}{\tilde J}{\tilde K}}$
 give a trivial deformation of the FDA
\eq{fda}. This means that the cocycle $e_{M{\tilde I}{\tilde J}{\tilde
    K}}$ is in fact a coboundary.
    According to the general construction of the free differential
    algebras, one can however expect that, if one enlarges the FDA by
    introducing
3-form potentials and the associated
curvatures,  a 4-form associated  to $e_{M{\tilde I}{\tilde J}{\tilde
    K}}$ could play a role. This
possibility will not be pursued here, but left for a future publication.



\section{The D=5 Rheonomic Lagrangian}
\label{rheonomic}

We write down explicitly here the rheonomic lagrangian, up to
4-fermion terms.
We recall that, in the rheonomic approach, the action is written as 
\begin{equation}
S=\int_{\cM_5} \cL
\end{equation}
where $\cM_5$ is a generic bosonic surface embedded in superspace.
$\cL$ is a 5-form written in terms of superfields without use of the Hodge-duality operator.
 It is then first-order in the kinetic terms, that is
auxiliary fields ${\bf F}^{\tilde I}_{ab}$, ${\bf X}^{\tilde I}_a$,
${\bf Q}^u_a$ are introduced and are then fixed in terms of the
physical field-strengths by solving their field equations. 
The space-time lagrangian is retrieved by restricting the rheonomic lagrangian along the space-time differentials $dx^\mu$, at zero fermionic coordinates $\Theta^A = d \Theta^A=0$.

With this approach, the field equations are valid all over superspace. The equations of motion on space-time are given by the field equations along the bosonic vielbein $V^a$ of superspace, while the field equations with at least one fermionic direction $\Psi^A$ yield the constraints on the supercurvatures and couplings.

The Lagrangian density up to 4-fermion terms can be written as:
\begin{equation}
\cL = \cL_{Grav} + \cL_{Kin} + \cL_{Pauli} + \cL_{CS} + \cL_{Tors} +
\cL_{gauge}
\label{eq:lagtot}
\end{equation}
where
\begin{eqnarray}
\cL_{Grav} &=& R^{ab} V^c V^d V^e \e_{abcde} -6 \i  \ol \Psi_A \G_{ab}
\rho^A V^a V^b \\
\cL_{Kin} &=& - \frac{3}{4} a_{{\tilde I}{\tilde J}} {\bf F}^{{\tilde
      I}}_{ab} \sx( F^{{\tilde J}} + 2 f^{{\tilde J}}_x \ol \Psi_A 
      \G_\ell \l^{xA} V^\ell \dx) \e^{ab}{}_{cde} V^c V^d V^e + \nn \\
&& +\left[ \frac{3}{4} g_{xy} {\bf X}^x_a \sx( D \varphi^y - \ol \Psi_A
\l^{yA} \dx)+g_{uv} {\bf Q}^u_a \sx( D q^v - \ol \Psi^A
\zeta^{\a} \U^v_{A{\a}} \dx)\right]  \e^a{}_{bcde} V^b V^c V^d V^e + \nn \\
&& - \frac{1}{10}\left[\frac34 \sx( g_{xy} {\bf X}^x_\ell {\bf X}^{y \ell}
- \frac{1}{2} a_{{\tilde I}{\tilde J}} {\bf F}^{{\tilde I}}_{fg} {\bf
      F}^{{\tilde J} fg}  \dx)+ g_{uv} {\bf Q}^u_\ell {\bf Q}^{v
      \ell}\right] \e_{abcde} V^a V^b V^c V^d V^e + \nn \\
&& +\left[\frac34 \i  g_{xy} \ol \l^{x}_A \G^a \na \l^{y A} +\frac
      12\i  \ol \zeta_{\a} \G^a \na \zeta^{\a}\right] \e_{abcde} V^b
      V^c V^d V^e \\ 
\cL_{Pauli} &=&  F^{{\tilde I}} \Bigl[ -\frac92 \i X_{\tilde I} \ol
  \Psi_A \G_a \Psi^A V^a -\frac92 h_{{{\tilde I}} x} \ol \Psi_A
  \G_{ab} \l^{xA}  V^a V^b  + \nn \\
&& \quad\quad- \frac 12 \i\left( \Phi_{{{\tilde I}} xy} \ol \l^{x}_A
  \G_{abc} \l^{yA} + 3 X_{\tilde I} \zeta_{\a}\G_{abc}\zeta^{\a}
  \right)V^a V^b V^c \Bigr] + \nn \\
&&\left( 3 g_{xy} D\varphi^x \ol \Psi_A \G_{abc} \l^{yA}+4
  \U_u{}^{A{\a}}Dq^u \ol \Psi_A \G_{abc} \zeta_{\a}\right)V^a V^b V^c
  \\ 
\cL_{Tors} &=& -3 \i  \mathcal{T}_aV^a\left( \ol \Psi_A \Psi^A
+\frac34 i   g_{xy} \ol \l^x_A \G_{bc} \l^{yA}  V^b V^c + \mez \ol
\zeta_\a \G_{bc} \zeta^\a V^b V^c \right) \label{eq:ltors} \\ 
\cL_{gauge} &=& \left(\frac32 g_{xy} W^{y\ AB} \ol \l_A^x \G^a \Psi_B
+ \N^{A}_{\a}\ol\Psi_A \G^a \zeta^{\a}\right) \e_{abcde} V^b V^c V^d
V^e + \nn \\ 
&& +6 \i  S^{AB} \ol \Psi_A \G_{abc} \Psi_B V^a V^b V^c
-\frac{1}{10}  V(\phi)  \e_{abcde} V^a V^b V^c V^d V^e + \nn \\
&& -\frac1{10} \left( 3\i M_{xy|AB} \ol \l^{xA} \l^{yB} +4\i
M^{A{\a}}_x\ol\zeta_{\a} \l^x_A +2 \i M^{{\a}{\b}}\ol\zeta_{\a}
\zeta_{\b}\right)\e_{abcde} V^a V^b V^c V^d V^e
\end{eqnarray}
The Chern-Simons Lagrangian can be written down in terms of
$A^{\tilde I}$ and $B_M$:
\begin{eqnarray}
\cL_{CS} &=& \a m^{MN} B_M \de B_N + s_{\tilde I}^{MN} B_M B_N A^{\tilde I} +
s_{{\tilde I}{\tilde J}}^I B_M A^{\tilde I} \de A^{\tilde J} +
s^I_{{\tilde I}{\tilde J}{\tilde K}} B_M A^{\tilde I} A^{\tilde J}
A^{\tilde K} + \nn \\
&& + \frac 34 t_{{\tilde I}{\tilde J}{\tilde K}} A^{\tilde I} \de
A^{\tilde J} \de A^{\tilde K} + r_{{\tilde I}{\tilde J}{\tilde K} |
  \tilde{L}} A^{\tilde I} A^{\tilde J} A^{\tilde K} \de A^{\tilde{L}}
+ r_{{\tilde I}{\tilde J}{\tilde K}\tilde{L}\tilde{M}} A^{\tilde I}
A^{\tilde J} A^{\tilde K} A^{\tilde{L}} A^{\tilde{M}}
\end{eqnarray}
where the gauge invariance of $\cL_{CS}$  implies:
\begin{eqnarray}
s_{\tilde I}^{MN} &=& \a m^{MP} T_{{\tilde I} P}{}^N \label{cs1}\\
s_{{\tilde I}{\tilde J}}^M &=& 2\a d_{{\tilde J}{\tilde I} N} m^{MN}
\label{css2}\\
s^M_{{\tilde I}{\tilde J}{\tilde K}} &=& \a m^{MN} d_{{\tilde{L}} [
    {\tilde I} | N} f_{{\tilde J}{\tilde K}]}{}^{\tilde{L}} \\
s_{({\tilde I}{\tilde J})}^M &=& \frac 14  t_{{\tilde I}{\tilde
    J}{\tilde K}} m^{{\tilde K}M} \label{csts}\\
r_{{\tilde I}{\tilde J}{\tilde K} | \tilde{L}} &=& \frac{1}{16}
    t_{{\tilde K}{\tilde{L}}\tilde{M}} f_{{\tilde I}{\tilde
    J}}{}^{\tilde{M}} + \frac{{\i}}{2} d_{{\tilde K} [ {\tilde I} | M}
    m^{MN} d_{{\tilde L} | {\tilde J}] N} \\
r_{{\tilde I}{\tilde J}{\tilde K}{\tilde{L}}{\tilde{M}}} &=&
    \frac{1}{80} t_{{\tilde I}{\tilde{N}}{\tilde{Q}}} f_{[{\tilde
    J}{\tilde K}}{}^{\tilde{N}}
    f_{{\tilde{L}}{\tilde{M}}]}{}^{\tilde{Q}} + \frac{{\a}}{5}
    d_{{\tilde I}[{\tilde J}M}m^{MN}d_{{\tilde{N}} | {\tilde K} J}
f_{{\tilde{L}}{\tilde{M}}]}{}^{\tilde{N}}\label{cs6}
\end{eqnarray}
Conditions \eq{cs1}-\eq{cs6}, required for gauge invariance of the
action, in particular imply that $t_{MNP} = 0$. To show this, let us
consider eq. \eq{csts} and mulitply it  by $m^{\tilde{I}N}
m^{\tilde{J}P}$:
\begin{equation}
\frac{1}{4} t_{{\tilde I}{\tilde J}{\tilde K}} m^{{\tilde I}M}
m^{{\tilde J}N} m^{{\tilde K}P} = s_{({\tilde I}{\tilde J})}^P
m^{{\tilde I}M} m^{{\tilde J}N}
\end{equation}
But due to eq. \eq{css2} this is related to the physical coupling $\hat{T}$:
\begin{equation}
\frac{1}{4} t_{{\tilde I}{\tilde J}{\tilde K}} m^{{\tilde I}M}
m^{{\tilde J}N} m^{{\tilde K}P} = 2 \a
d_{({\tilde I}{\tilde J})Q} m^{PQ} m^{{\tilde I}M} m^{{\tilde J}N} =
\a \hat{T}_{\tilde{I}Q}{}^N m^{{\tilde I}M} m^{PQ} \label{tmnp0}
\end{equation}
This last term vanishes due to eq. \eq{closure2}, so that, since we generally take $m^{MN}$ invertible, it gives:
\begin{equation}
t_{MNP}=0\,. \label{tmnp0}
\end{equation}

As a final remark, let us observe that, given \eq{cs1} - \eq{cs6}, all
the Chern-Simons Lagrangian \eq{cs} contains a free multiplicative
parameter, $\alpha$. However, recalling the discussion at the end of
section \ref{generalities}, the theory still has the scale invariance
\eq{scaleinv}, which may be used to fix the parameter $\a$ at our
wish. We set $\a = \frac{9}{4}$. This finally gives eq. \eq{csbis}.

\section{The four-fermions Lagrangian}\label{4f}

The 4-fermions contributions to the Lagrangian, from
\cite{Gunaydin:1983bi}, with our notations reads: 
\begin{eqnarray}
\cL_{4f}&=&\Bigl\{ -\frac 1{16} D_w T_{xyz} \bar \l^x_A
\G_{\m\n}\l^{yA}\bar \l^z_B  \G^{\m\n}\l^{wB} + \mez
R_{xyzw}\left(\bar \l^x_A \l^{yA}\bar \l^z_B  \l^{wB} +\bar \l^x_A
\G_{\m}\l^{yA}\bar \l^z_B  \G^{\m}\l^{wB}\right) +\nn\\
&&-\frac 34 g_{xy} g_{zw}\left(\bar \l^x_A \l^{y}_B\bar \l^{zA}
\l^{wB} +\frac 12 \bar \l^x_A \G_{\m}\l^{y}_B\bar \l^{zA}
\G^{\m}\l^{wB} - \frac 3{16}\bar \l^x_A \G_{\m\n}\l^{y}_B\bar \l^{zA}
\G^{\m\n}\l^{wB}\right)\Bigr\} _{4\l}+\nn\\
&&\Bigl\{2\i T_{xyz} \left(\bar \l^x_A \Psi_{\mu B}\bar \l^{yA}
\G^{\m}\l^{zB} -\frac 12 \bar \l^x_A \G^{\m}\Psi_{\mu B}\bar \l^{yA}
\l^{zB}\right)\Bigr\}_{3\l}+\nn\\
&& \Bigl\{\frac 3{16} g_{xy}\Bigl[\bar \Psi_{\m A} \Psi_{\nu B }\bar
  \l^{xA}  \left(3 g^{\m\n}+\frac 12\G^{\m\n}\right)\l^{yB} -\bar
  \Psi_{\m A} \G_\rho\Psi_{\nu B }\bar \l^{xA}  \left(3
  g^{\m\n}\G^\rho +2g^{\m\rho}\G^\n
  +\frac12\G^{\m\n\rho}\right)\l^{yB} +\nn\\
&&-\frac 12 \bar \Psi_{\m A} \G_{\rho\s}\Psi_{\nu B }\bar \l^{xA}
  \left(2 g^{\m\rho}g^{\n\s} -4g^{\m\rho}\G^{\n\s}
  -2g^{\m\n}\G^{\rho\s}
  +\frac12\G^{\m\n\rho\s}\right)\l^{yB}\Bigr]\Bigr\}_{2\l} +\nn\\
  &&+\Bigl\{\frac 3{16}\bar\Psi_{\m A}\Psi_\n^A \bar\Psi^\m_B \Psi^{\n B} +
  \frac 18 \bar\Psi_{\m A}\G^{\m\n\r\s}\Psi_\n^A \bar\Psi_{\r B} \Psi^{B}_\s +
  \frac 12 \bar\Psi_{\n A}\G^{\n}\Psi_\m^A \bar\Psi_{\r B}\G^\r
  \Psi^{\m B} +\nn\\
  &&+ \frac 18 \bar\Psi_{\m A}\G_\r\Psi_\n^A \bar\Psi^\m_{ B} \G^\r\Psi^{\n B}
  -\frac 12 \bar\Psi_{(\n A}\G_{\r)}\Psi_\m^A \bar\Psi^\n_{ B} \G^\r\Psi^{\m B}
 \Bigr\}_{4\Psi}+\nn\\
&&+  \ol \Psi_{A|\m} \zeta_{\a} \ol \Psi^A_\n ( g^{\m\n} +
\G^{\m\n} ) \zeta^{\a}  + \frac{1}{16} \ol \zeta_{\a} \G_{\m\n} \zeta^{\a}
\ol \zeta_{\b} \G^{\m\n} \zeta^{\b} + \\
&& + \frac{3}{16} g_{xy} \ol \zeta_{\a} \G_{\m\n\rho} \zeta^{\a} \ol
\l^x_A \G^{\m\n\rho} \l^{yA} - \qu \Omega^{{\a}{\b}{\g}{\d}} (5 \ol \zeta_{\a}
\zeta_{\b} \ol \zeta_{\g} \zeta_{\d} - \ol \zeta_{\a} \G_\m \zeta_{\b} \ol
\zeta_{\g} \G^\m \zeta_{\d} )
\end{eqnarray}
where $\Omega^{{\a}{\b}{\g}{\d}} = \cR^{{\a}{\b}}{}_{uv} \U^{u|{\g}A}
\U^{v|{\d}B}\e_{AB}$.

\section{Useful relations with $\G$-matrices and Fierz identities}

\begin{eqnarray}
\e^{a_1 \dots a_p b_1 \dots b_q}\e_{a_1 \dots a_p c_1 \dots c_q}
&=&p! q! \d^{ b_1 \dots b_q}_{ c_1 \dots c_q}\,,\qquad (p+q=5)\\
\G^{abcd} &=& \e^{abcde} \G_e\\
\G^{abc} &=& - \mez \e^{abcde} \G_{de}\\
\G_a \G^{bc} &=& \G_a{}^{bc} + 2 \d_a^{[b} \G^{c]}\\
\G^{bc} \G_a &=& \G_a{}^{bc} - 2 \d_a^{[b} \G^{c]}\\
\G^{ab} \G_{cd} &=& \G^{ab}{}_{cd} - 4 \d^{[a}_{[c} \G^{b]}{}_{d]} -
2
\d^{ab}_{cd} \\
\G^{[a}\G_{cd} \G^{b]}&=& \G^{abcd} + 2 \delta^{ab}_{cd} \\
\G^{a}\G_{a b_1 \dots b_p}&=&(5-p) \G_{b_1 \dots b_p}\,,\qquad 0\leq
p\leq 4\\
\G^{ab}\G_{abc_1 \dots c_p}&=&-(5-p)(4-p) \G_{ c_1 \dots
  c_p}\,,\qquad 0\leq p\leq 3\\
\G^{a}\G_{c}\G_{ab}&=&2\G_{bc} - 4\d_{bc}\\
\G^{a}\G_{cd}\G_{a}&=&\G_{cd}\\
\G^{a}\G_{cd}\G_{ab}&=&-4\eta_{b[c}\G_{d]}= \G_{ab}\G_{cd}\G^{a}\\
\G^{a}\G_{c}\G_{a}&=&-3\G_{c}\\
\G^{ab}\G_{c}\G_{ab}&=&-4\G_{c}\\
\G^{ab}\G_{cd}\G_{ab}&=&4\G_{cd}
\end{eqnarray}

Recalling that
\begin{equation}
\Psi_A \equiv \epsilon_{AB}\Psi^B\,; \qquad \Psi^A =- \epsilon^{AB}\Psi_B
\end{equation}
and that the currents of spinor one-forms have  the symmetry properties
\begin{eqnarray}
\ol\Psi_A \Psi_B &=& - \ol\Psi_B\Psi_A\quad  \left(= -\frac 12\e_{AB}
\ol\Psi_C \Psi^C\right)\\
\ol\Psi_A \G^a\Psi_B &=& - \ol\Psi_B\G^a\Psi_A\\
\ol\Psi_A \G^{ab}\Psi_B &=&  \ol\Psi_B\G^{ab}\Psi_A
\end{eqnarray}
the following Fierz-identities follow:
\begin{eqnarray}
 {\Psi}_A \wedge  {\ol{\Psi}}_B &=&\frac{1}{4} \sx( {\ol{\Psi}}_B
 {\Psi}_A+\G_a
 {\ol{\Psi}}_B \G^a  {\Psi}_A
\dx) - \frac{1}{8} \G_{ab}
 {\ol{\Psi}}_B \G^{ab}  {\Psi}^A  \label{eq:fierz5-1}\\
  {\Psi}_A \wedge  {\ol{\Psi}}_B\wedge  \Psi_C  &\equiv&-\frac 12
 \epsilon_{BC} \Xi_A
\label{eq:fierz5-2}\\
  {\Psi}_A \wedge  {\ol{\Psi}}_B\wedge \G^a \Psi_C  &=&-\frac 12
  \epsilon_{BC}\sx( \Xi^a_A  + \frac 15 \G^a \Xi_A\dx), \qquad \G_a
  \Xi^a_A =0\label{eq:fierz5-3}\\
  {\Psi}_A \wedge  {\ol{\Psi}}_B\wedge \G^{ab} \Psi_C &=&
  \Xi^{ab}_{(ABC)} -\frac 23 \e_{A(B} \G^{[a} \Xi^{b]}_{C)} +\frac 15
  \e_{A(B} \G^{ab} \Xi_{C)},\qquad \G_a \Xi^{ab}_{(ABC)}= 0
  \label{eq:fierz5-4}
\end{eqnarray}
so that
\begin{eqnarray}
 \G^a  {\Psi}^A \wedge  {\ol{\Psi}}_B\wedge \G_a \Psi^B  &=&
 {\Psi}^A   \wedge  {\ol{\Psi}}_B\wedge
\Psi^B\label{eq:summedfierz5-1}\\
\G^{ab}  {\Psi}^A \wedge  {\ol{\Psi}}_B\wedge \G_{ab} \Psi_C &=& -4
\,\delta^A_{(B} {\Psi}_{C)}   \wedge  {\ol{\Psi}}_L\wedge
\Psi^L\label{eq:summedfierz5-2}
\end{eqnarray}

\section{Matching of notations}
\label{rosetta}

We have collected in the first table the differences in notation
between \cite{Gunaydin:1983bi} (GST) and the present paper (ADS) for
the vector and tensor multiplet sector, and in the second the differences
with \cite{Ceresole:2000jd} (CD) for the hypermultiplet sector.

\begin{table}[h]
\begin{center}
\begin{tabular}[t]{|l|l|l|l|l|l|l|l|l|l|}
\hline
GST & $\eta_{ab}$ & $\G^a$ & $\G_a$ & $\epsilon^{AB}$ & $h^{\tilde I}$
& $h_{\tilde I}$ & $F^{\tilde I}_{ab}$ & $\phi^x$ & $\psi^{A}$ \\
\hline
ADS & $-\eta_{ab}$ & $\i\G^a$ & $-\i\G_a$ & $-\epsilon^{AB}$ &
$\frac{4}{\sqrt{6}} X^{\tilde I}$ & $\frac{\sqrt{6}}{4} X_{\tilde I}$
& $2 F^{\tilde I}_{ab}$ & $\phi^x$ & $ \sqrt{2} \psi^{A}$ \\
\hline 
\hline
\hline
GST & $h^{\tilde I}_x$ & $h_{\tilde I}^x$ & $\dot a_{{\tilde 
I}{\tilde J}}$ & $g_{xy}$ & $T^x{}_{yz}$ & $\phi_{{\tilde I} xy}$ & 
$C_{{\tilde I} {\tilde J}{\tilde K}}$ & $K^x_{\ yzt}$ &
$\l^{xA}$ \\
\hline
ADS & $-2 f^{\tilde I}_x$ & $-\mez g^x_{\tilde I}$ &
$\frac{3}{8}\, a_{{\tilde I}{\tilde J}}$ & $\frac{3}{2}\,g_{xy}$ & 
$-\sqrt{\frac 32} \,T^x{}_{yz} $ & $\frac{1}{4}\,\phi_{{\tilde I} xy}$
& ${\sqrt\frac{27}{8}}\, t_{{\tilde I}{\tilde J}{\tilde K}}$ &
$R^x_{\ yzt}$ & $ - \i \sqrt{2} \l^{xA}$ \\
\hline
\end{tabular}
\end{center}
\label{GST/ADS}
\end{table}%

\begin{table}[h]
\begin{center}
\begin{tabular}[t]{|l|l|l|l|l|l|l|l|l|}
\hline
CD & $C_{\a\b}$ & $q^u$ & $\zeta^\a$ & $\U_u{}^{A\a}$ & $g_{uv}$ &
$k^{u}_{\tilde{I}}$ & $\N^{A\a}$ & $\M^{A\a}_x$ \\
\hline
ADS & $- C_{\a\b}$ & $q^u$ & $ - \i \zeta^\a$ & $ - {\sqrt{2}}
\U_u{}^{A\a}$ & $2g_{uv}$ & $k^u_{\tilde{I}}$ & $-\frac1{\sqrt{2}}
\N^{A\a}$ & $-\frac1{\sqrt{2}} \M^{A\a}_x$ \\ 
\hline
\end{tabular}
\end{center}
\label{CD/ADS}
\end{table}%

\end{document}